\documentclass[aps, prb, amsmath, twocolumn, 10pt]{revtex4-2}

\usepackage{threeparttable}
\usepackage{booktabs}
\usepackage{graphicx}
\usepackage{braket}
\usepackage{upgreek}

\usepackage{xcolor}
\newcommand{\col}{black}

\begin{document}

\title{Light-hole spin confined in germanium}

\author{Patrick Del Vecchio}
\affiliation{Department of Engineering Physics, \'Ecole Polytechnique de Montr\'eal, Montr\'eal, C.P. 6079, Succ. Centre-Ville, Montr\'eal, Qu\'ebec, Canada H3C 3A7}
\author{Oussama Moutanabbir}
\email{oussama.moutanabbir@polymtl.ca}
\affiliation{Department of Engineering Physics, \'Ecole Polytechnique de Montr\'eal, Montr\'eal, C.P. 6079, Succ. Centre-Ville, Montr\'eal, Qu\'ebec, Canada H3C 3A7}

\begin{abstract}
The selective confinement of light holes (LHs) in a tensile-strained germanium (Ge) quantum well is studied by mapping the electronic structure of Ge$_{1-x}$Sn$_x$/Ge/Ge$_{1-x}$Sn$_x$ heterostructures as a  function of Sn content, residual strain, and Ge well thickness. It is shown that above $12\,\text{at.}\%$ Sn and below $0.4\%$ residual compressive strain in the barriers, the tensile strain in Ge becomes sufficiently large to yield a valence band edge with LH-like character, thus forming a quasi two-dimensional LH gas in Ge. The LH ground state has a larger in-plane effective mass than that of heavy holes (HHs) in Si$_{1-y}$Ge$_y$/Ge/Si$_{1-y}$Ge$_y$ quantum wells. Moreover, LHs in optimal Ge$_{1-x}$Sn$_x$/Ge/Ge$_{1-x}$Sn$_x$ heterostructures are found to exhibit a strong $g$-tensor anisotropy, with the in-plane component one order of magnitude larger than that of HHs in typical planar systems. Two of three structure-inversion-asymmetry Rashba parameters, both of which are critical in electric-dipole-spin-resonance experiments, are effectively 10 times the size of the cubic Rashba parameter in HH quantum wells. In the regime of LH selective confinement, every layer of the heterostructure is of direct bandgap, which can be relevant for efficient optical photon-spin qubit interfaces. This work discusses the broad landscape of the characteristics of LH spin confined in Ge to guide the design and implementation of LH spin-based devices.
\end{abstract}

\maketitle

\section{Introduction}

Hole spins in group-IV planar gated quantum dots are promising candidates for robust and scalable qubits~\cite{Fang_2023,Lawrie2023,Jirovec2021,Hendrickx2021,Hendrickx2020_1,Hendrickx2020_2,Lawrie2020}. Developing these qubits has been so far exclusively based on heavy-hole (HH) states, as the materials currently used are restricted to compressively strained germanium (Ge) heterostructures. Interestingly, the advent of the germanium/germanium-tin (Ge/Ge$_{1-x}$Sn$_{x}$) material system provides an additional degree of freedom to tailor the valence band character in tensile-strained Ge, thus paving the way to implement silicon-compatible platforms for light-hole (LH) spin qubits~\cite{Assali2022,DelVecchio2023}. These LH qubits share many of the advantages benefiting the HH ones but also bring about other attractive characteristics pertaining to LHs and Ge/Ge$_{1-x}$Sn$_{x}$ heterostructures. These include a strong Rashba-type spin-orbit interaction (SOI)~\cite{DelVecchio2023} and an efficient coupling with proximity-induced superconductivity~\cite{Moghaddam2014} in addition to the bandgap directness~\cite{Moutanabbir2021} relevant to coupling with optical photons. These characteristics can expand the functionalities of hole spin qubits by facilitating the implementation of hybrid superconductor-semiconductor devices and photon-spin interfaces. 

Although Ge$_{1-x}$Sn$_{x}$ semiconductors have been the subject of extensive studies in recent years, research in this area has mainly been focused on integrated photonics and optoelectronics leveraging Ge$_{1-x}$Sn$_{x}$ strain- and composition-dependent bandgap \cite{Moutanabbir2021} leaving their spin properties practically unexplored~\cite{Tai2021,Ferrari2023,Fettu2023,DelVecchio2023}. As a matter of fact, studies on hole spin in Ge/Ge$_{1-x}$Sn$_{x}$ are still conspicuously missing in the literature.  Herein, this work addresses the dynamics of LH spin confined in Ge$_{1-x}$Sn$_{x}$/Ge/Ge$_{1-x}$Sn$_{x}$ heterostructures and elucidates the key parameters affecting its behavior as a function of strain, well thickness, Sn content, and  magnetic field orientation. First, the article describes and discusses the electronic structure of tensile-strained Ge on Ge$_{1-x}$Sn$_{x}$. Second, the parameters defining the band alignment of the Ge$_{1-x}$Sn$_x$/Ge/Ge$_{1-x}$Sn$_x$ quantum well are introduced and the criteria for LH confinement in Ge are established. The third section outlines the Hamiltonian of the in-plane motion of LHs for out-of-plane and in-plane magnetic fields yielding LH parameters such as the effective mass, the $g$-tensor, and the Rashba-SOI parameters. LH-HH mixing within the LH ground state is \textcolor{\col}{investigated} in the fourth section. It is important to note that the focus here is on LH properties in the planar system without the effects of electrostatic in-plane confinement introduced in quantum dot systems.~\cite{DelVecchio2023}

\section{LH quantum well in Ge$_{1-x}$Sn$_{x}$/Ge/Ge$_{1-x}$Sn$_{x}$}

\subsection{Ge$_{1-x}$Sn$_{x}$/Ge/Ge$_{1-x}$Sn$_{x}$ heterostructure}

\begin{figure}[t]
    \centering
    \includegraphics[width=\linewidth]{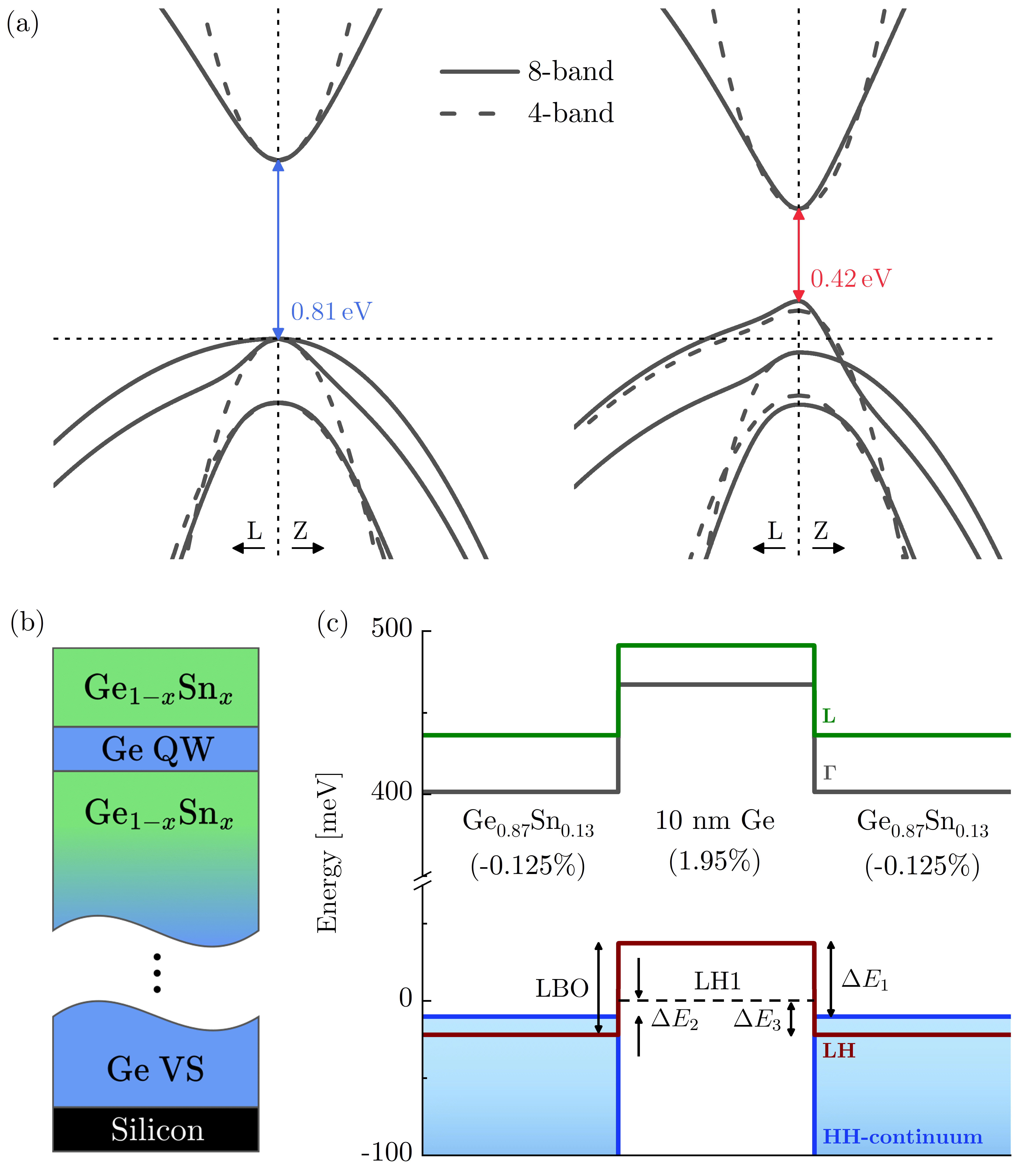}
    \caption{(a) Bulk dispersion near the $\Gamma$ point ($\|\mathbf{k}\| \leq 2.5\,\text{nm}^{-1}$ in both directions) of relaxed Ge (left) and tensile-strained Ge (right) computed by the 8-band (full lines) and the 4-band (dashed lines) $k\cdot p$ frameworks. The nonzero components of the strain tensor are $\varepsilon_{xx}=\varepsilon_{yy} = 2\%$ and $\varepsilon_{zz} = -2(c_{12}/c_{11})\varepsilon_{xx} \approx -1.3\%$. The conduction band was obtained from an effective mass approximation in the 4-band computations. (b) Schematic illustration of the proposed Ge$_{1-x}$Sn$_{x}$/Ge/Ge$_{1-x}$Sn$_{x}$ heterostructure. (c) Band alignment profile of a Ge$_{1-x}$Sn$_{x}$/Ge/Ge$_{1-x}$Sn$_{x}$ quantum well for selected barrier parameters. In this case, $\Delta E_{1,2,3}$ and LBO are positive. $\varepsilon_{xx}$ in each layer is written in parentheses.}
    \label{fig:stack}
\end{figure}

Before delving into the details of the electronic structure of Ge/Ge$_{1-x}$Sn$_{x}$, it is instructive to examine the strain-related behavior of bulk Ge. Figure~\ref{fig:stack}(a) outlines the effect of tensile strain on the band structure of bulk Ge calculated by 8-band (full lines) and 4-band (dashed lines) $k\cdot p$ theory. Here, the calculations assume a bi-isotropic biaxial strain in the $(001)$-plane without any shear deformations, which is expected for an ideal $[001]$-oriented epitaxial growth~\textcolor{\col}{\cite{VandeWalle1989,Assali2022}}. Under these conditions, the fourfold degeneracy of the VB edge at the $\Gamma$ point is lifted yielding two spin-degenerate LH and HH bands. In the case of tensile strain, the VB edge is LH-like, whereas in the broadly studied compressively strained Ge, the VB edge is of the HH type. As discussed in the following, there is a threshold of tensile strain beyond which it becomes possible to control and selectively manipulate spin 1/2 LHs instead of spin 3/2 HHs.

The proposed heterostructure consists of a Ge$_{1-x}$Sn$_{x}$/Ge/Ge$_{1-x}$Sn$_{x}$ quantum well grown on silicon~\cite{Assali2022}, as illustrated in Fig.~\ref{fig:stack}(b). Thick Ge$_{1-x}$Sn$_{x}$ buffer layers with an increasing Sn content from the Ge virtual substrate (VS) up to the bottom Ge$_{1-x}$Sn$_x$ barrier prevents the propagation of defects and dislocations near the Ge quantum well~\cite{Assali2022}. \textcolor{\col}{The lattice mismatch between Ge$_{1-x}$Sn$_{x}$ and Ge (with Sn contents above ${\sim}10\,\text{at.}\,\%$) is leveraged to achieve high tensile strain in the coherently grown Ge quantum well}. A top Ge$_{1-x}$Sn$_{x}$ barrier is then grown on the tensile-strained Ge layer. As shown in the following, \textcolor{\col}{careful engineering of the lattice strain and Sn content leads to a direct bandgap Ge LH quantum well.}

\subsection{Band alignment and LH confinement}

The energy band alignment of the Ge$_{1-x}$Sn$_{x}$/Ge/Ge$_{1-x}$Sn$_{x}$ heterostructure is computed within the assumption that the in-plane lattice constant $a_\parallel$ is the same in each layer, which is consistent with \textcolor{\col}{a} pseudomorphic epitaxial growth. The in-plane components $\varepsilon_{xx}$ and $\varepsilon_{yy}$ of the strain tensor in a material $i$ are given by $\varepsilon_{xx}=\varepsilon_{yy}=a_\parallel/a_0^i-1$, where $a_0^i$ is the lattice constant of the fully relaxed material $i$. The out-of-plane strain component $\varepsilon_{zz}$ then follows immediately from the relation $\varepsilon_{zz} = -2(c_{12}/c_{11})\varepsilon_{xx}$, where $c_{11}$ and $c_{12}$ are the material elastic constants (see Table \ref{tb:parameters}) and $\varepsilon_{kl}=0$ if $k\neq l$~\textcolor{\col}{\cite{VandeWalle1989}}. Three parameters determine completely the band alignment of the  Ge$_{1-x}$Sn$_{x}$/Ge/Ge$_{1-x}$Sn$_{x}$ heterostructure: the Sn content $x$, the in-plane strain $\varepsilon_\text{BR}$ in the barriers, and the in-plane strain in the Ge well $\varepsilon_\text{W}$. \textcolor{\col}{Since t}he latter can be counted for from the condition that $a_\parallel$ is constant along the growth direction $z$, \textcolor{\col}{only the barrier composition and strain ($x$, $\varepsilon_\text{BR}$) are required to evaluate the band alignment.}

A typical band alignment is displayed in Fig. \ref{fig:stack}(c) for $x=0.13$ and $\varepsilon_\text{BR}=-0.125\%$. \textcolor{\col}{In this instance}, HHs and electrons are pushed away from Ge and form a continuum of states in the Ge$_{0.87}$Sn$_{0.13}$ barriers. \textcolor{\col}{Meanwhile, LHs are selectively confined in Ge, thereby forming a LH quantum well. Here, the combination of large strain in Ge ($\varepsilon_\text{W} = 1.95\%$) with small $\varepsilon_\text{BR}$} pushes the LH ground state (LH1) above the HH continuum, \textcolor{\col}{leading to the possibility}, at very low hole density, to populate only LH1 and thus \textcolor{\col}{to} create a pure quasi two-dimensional LH gas in Ge. \textcolor{\col}{Such} system is achievable only in a specific region of the parameter space ($x$, $\varepsilon_\text{BR}$), depending on four energy offsets [see Fig. \ref{fig:stack}(c)]:

\begin{align}
  \Delta E_1 &= \max(\text{LH}) - \max(\text{HH}), \\
  \Delta E_2 &= E_{\text{LH}1} - \max(\text{HH}), \\
  \Delta E_3 &= E_{\text{LH}1} - \min(\text{LH}), \\
  \begin{split}
  \text{LBO} &= \max(\text{LH}) - \min(\text{LH}) \\
  &= \Delta E_1 - \Delta E_2 + \Delta E_3.
  \end{split}
\end{align}

\noindent Here, $\max(\text{LH})$ and $\min(\text{LH})$ are the energies at the bottom and top of the LH quantum well, respectively, and $\max(\text{HH})$ is the energy at the edge of the HH continuum. The zero energy point is placed on the ground LH subband (i.e., $E_\text{LH1}=0$). Band offsets $\Delta E_1$ and LBO (LH band offset) do not depend on the well thickness $w$.

\begin{figure}[t]
    \centering
    \includegraphics[width=\linewidth]{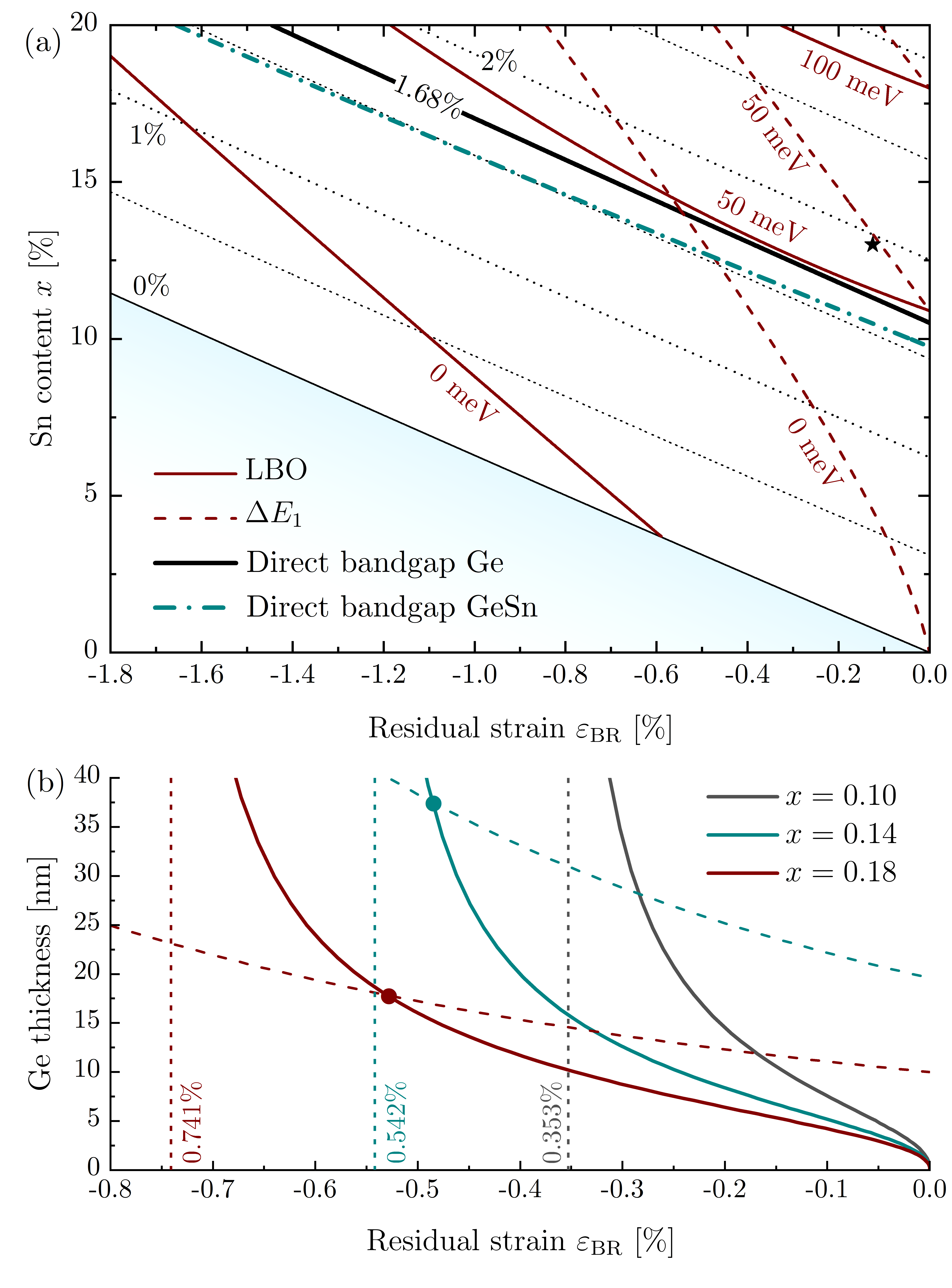}
    \caption{(a) Two-dimensional contour map of band offsets $\Delta E_1$ and LBO, strain in Ge, and bandgap directness as a function of $x$ and $\varepsilon_\text{BR}$ at $T=300\,\text{K}$. Contour lines for LBO and $\Delta E_1$ are given at $0$, $50$, and $100\,\text{meV}$. The star indicates the system from Fig. \ref{fig:stack}(c). (b) Minimal well thickness $w_0$ (solid lines) required for a LH-like valence band edge as a function of $\varepsilon_\text{BR}$ at fixed Sn content in the barriers. Dashed lines are the Ge critical growth thickness $h_c$ estimated from the People and Bean relation. Solid circles indicate where $w_0 = h_c$ for a given $x$. For $x = 0.10$, $h_c\approx 46\,\text{nm}$ at $\varepsilon_\text{BR} = 0$.}
    \label{fig:conf}
\end{figure}

Figure \ref{fig:conf}(a) presents a two-dimensional map of band offsets LBO and $\Delta E_1$ with $x$ and $\varepsilon_{\text{BR}}$ as independent parameters. The corresponding strain in the Ge well $\varepsilon_\text{W}$ is also shown (black dotted lines) only for the tensile strain regime ($\varepsilon_\text{W}>0$). \textcolor{\col}{The Ge indirect-to-direct transition occurs at $\varepsilon_\text{W} = 1.68\%$ (solid black line) according to the parameterization described in Appendix \ref{sec:params}}. Similarly, the Ge$_{1-x}$Sn$_{x}$ barriers exhibit bandgap directness above the dashed-dot blue line. Constant LBO are indicated by the solid red curves, where $\text{LBO}=0$ corresponds to a completely flat LH profile along the growth direction. Finally, dashed red curves indicate constant $\Delta E_1$, where $\Delta E_1 = 0$ corresponds to the LH band edge in Ge sitting at the same energy as the HH band edge in the barriers. As discussed in the following, a \textcolor{\col}{large and positive} $\Delta E_1$ allows LH1 to emerge from the continuum for sufficiently thick quantum wells.

\textcolor{\col}{LHs are confined in Ge if and only if} $\text{LBO}>0$. However, depending on $\Delta E_1$ and the well thickness $w$, \textcolor{\col}{these confined LHs could be situated within} the HH continuum (negative $\Delta E_2$). In the region $\Delta E_1 \leq 0$, LH\textcolor{\col}{s} can never emerge from the continuum since the bottom of the LH well is below the HH band edge. However, if $\Delta E_1 > 0$, there is a minimal \textcolor{\col}{QW} thickness $w_0$ for which LH1 emerges from the continuum. \textcolor{\col}{This lower bound} depends on both Sn content and lattice strain (i.e., $x$ and $\varepsilon_\text{BR}$) in the barriers and is plotted in Fig. \ref{fig:conf}(b) (solid lines). At $w = w_0$, the energy required for \textcolor{\col}{LH1} to escape the well is exactly $\text{LBO} - \Delta E_1$ (i.e., the strain-induced HH-LH splitting in the barriers). \textcolor{\col}{$w_0 \to \infty$ on the $\Delta E_1 = 0$ curve}, whereas $w_0 = 0$ if the barriers are fully relaxed. \textcolor{\col}{In addition to} the lower bound $w_0 < w$, \textcolor{\col}{the maximum strain energy that the Ge QW can accommodate also introduces an upper bound on $w$}. This upper bound is given by the Ge critical growth thickness $h_c$, beyond which misfit dislocations start to appear at the interfaces and tensile strain in Ge is strongly suppressed. The dashed lines in Fig.~\ref{fig:conf}(b) show an estimation of $h_c$ based on the People and Bean formula~\cite{People1985,People1986E}. Here, this model is applied \textcolor{\col}{for} the critical thickness of a Ge layer \textcolor{\col}{with} equilibrium lattice constant $a_0^\text{Ge}\equiv a_0(0)$ pseudomorphically grown on Ge$_{1-x}$Sn$_x$ \textcolor{\col}{with} lattice constant $a_0(x)\left(1 + \varepsilon_\text{BR}\right)$.

\section{LH spin properties}

\subsection{Effective masses and spin parameters}

In this section, the LH subband effective mass, the out-of-plane and in-plane $g$ factors, and the Rashba parameters are computed as a function of Sn content in the Ge$_{1-x}$Sn$_{x}$ barriers ($x$) and well thickness $w$. These parameters give important information on how LHs move in the plane and how they respond to magnetic fields. Despite being intrinsic to a given subband (here, the focus is given to the lowest subband LH1), there is generally a significant influence from neighboring levels through inter-subband couplings. Moreover, the wavefunction spread across interfaces can also influence subband parameters. In heterostructures such as Ge/Si$_{1-y}$Ge$_{y}$ quantum wells,  where both types of holes are confined into the same layer and the band offsets are large, intersubband coupling is significant to an energy scale comprising only a few tens of subbands due to quantization effects. For instance, it is often a reasonable approximation to include the coupling to only 1 or 2 LH subbands for a HH ground state~\cite{Wang2021}, or around 50 LH subbands when band offsets are taken into account~\cite{Wang2022arxiv}. In Ge/Ge$_{1-x}$Sn$_{x}$ quantum wells, intersubband couplings must include ${\sim} 10^2$ subbands due to the neighboring continuum. Moreover, small LBOs leading to a sizable spread of LH1 into the barriers require an accurate description of the subband envelopes. To address these effects, a numerical approach is employed instead of a variational method for the envelope problem~\cite{DelVecchio2023}. Moreover, due to the nearby continuum, the coupling from the $200$ closest subbands to LH1 are taken into account. \textcolor{\col}{Our implementation of $k\cdot p$ theory and how strain is incorporated into the model is described in Appendix \ref{sec:model}.}

From the point of view of 8-band $k\cdot p$ theory, a subband such as LH1 always consists of the superposition of spin $1/2$ CB electron, LH, and split-off (SO) hole envelopes. At $\mathbf{k}_\parallel = 0$, this can be written as

\begin{equation}\label{eta}
    \Ket{\eta,\sigma} = \Ket{\frac{1}{2},\frac{\sigma}{2}}_c\Ket{c} + \Ket{\frac{3}{2},\frac{\sigma}{2}}\Ket{\ell} + \sigma\Ket{\frac{1}{2},\frac{\sigma}{2}}\Ket{s},
\end{equation}

\noindent where $\sigma=\pm 1$ is a pseudo-spin quantum number and $\Braket{z|c,\ell,s} = \psi_{c,\ell,s}(z)$ are the CB, LH, and SO envelope functions, respectively. \textcolor{\col}{To avoid any confusion}, ``LH'' subbands (e.g., LH1) are \textcolor{\col}{designated} as $\eta$ subbands to \textcolor{\col}{distinguish} them from their LH envelope component. The kets $\Ket{\frac{3}{2},\frac{\sigma}{2}}\equiv \ket{j,m}$ and so on are the bulk Bloch states. The additional contribution from HHs away from $\mathbf{k}_\parallel = 0$ \textcolor{\col}{is} investigated in the next section. An $\eta$ subband is normalized according to

\begin{equation}
    1 = \sum_{\tau = \{c,\ell,s\}}{\Braket{\tau|\tau}},
\end{equation}

\noindent with $\Braket{\ell|\ell} > \Braket{c|c}$ and $\Braket{\ell|\ell} > \Braket{s|s}$ for a level such as LH1. For instance, the SO contribution $\Braket{s|s}$ in LH1 is typically smaller than $10\%$ for the range of barrier and well parameters considered here, but it plays an important role in the effective mass calculations, as discussed in the following. The CB contribution $\Braket{c|c}$ in LH1 is around $5\%$, where the envelopes $\Ket{c}$ are antisymmetric with respect to the center of the well and have their maximal amplitude near the interfaces.

For an out-of-plane magnetic field $\mathbf{B} = B\mathbf{e}_z$, the in-plane motion of $\Ket{\eta,\sigma}$ is described by an effective two-dimensional Hamiltonian (see Appendix~\ref{sec:pert})

\begin{equation}\label{eff_par}
  \begin{split}
    H_{\text{eff}}^\perp &= \alpha_0\gamma K_\parallel^2 + \frac{\alpha_0}{\lambda^2}\frac{g_\perp}{2}\sigma_z \\
    &+ i\beta_1\left(K_-\sigma_+ - \text{h.c.}\right) - i\beta_2\left(K_+^3\sigma_+ - \text{h.c.}\right) \\
    &+ i\beta_3\left(K_-K_+K_-\sigma_+ - \text{h.c.}\right),
  \end{split}
\end{equation}

\noindent where $\alpha_0 = \hbar^2/(2m_0)$ with $m_0$ the free electron mass, $\mathbf{K} = \mathbf{k} + e\mathbf{A}/\hbar$ is the mechanical wavevector, $\mathbf{A} = B/2(-y\mathbf{e}_x+x\mathbf{e}_y)$ is the vector potential such that $\mathbf{B} = \nabla\times\mathbf{A}$, and $\mathbf{k}\to -i\nabla$ is the canonical wavevector. Notably, $K_\parallel^2 = \{K_-,K_+\}/2 = K_x^2+K_y^2$, $1/\lambda^2 = [K_-,K_+]/2 = eB/\hbar$, and $K_\pm = K_x \pm iK_y$. The effective parameters in \eqref{eff_par} are the following: $\gamma = m_0/m^*$ is the in-plane inverse effective mass, $g_\perp$ is the out-of-plane $g$ factor, and $\beta_{1,2,3}$ represent the three types of Rashba splittings. The first is linear in $K$, whereas the last two are cubic in $K$. Rashba parameters arise from space inversion asymmetry. When a small DC electric field $\mathbf{E} = E_z\mathbf{e}_z$ is applied to an otherwise symmetric well, all $\beta$ parameters behave linearly with $E_z$ and involve only odd powers of the field:

\begin{equation}
    \beta_i = \alpha_iE_z + O\left(E_z^3\right),
\end{equation}

\noindent where $i=1,2,3$. For an in-plane magnetic field $\mathbf{B} = B(\mathbf{e}_x\cos\phi + \mathbf{e}_y\sin\phi)$ with vector potential $\mathbf{A} = B(\mathbf{e}_x\sin\phi - \mathbf{e}_y\cos\phi)z$, the in-plane motion is instead described by

\begin{equation}\label{eff_in}
  \begin{split}
    H_{\text{eff}}^\parallel &= \alpha_0\gamma k_\parallel^2 + \frac{\alpha_0}{\lambda^2}\frac{g_\parallel}{2}\left(e^{-i\phi}\sigma_+ + \text{h.c.}\right) \\
    &+ i\beta_1\left(k_-\sigma_+ - \text{h.c.}\right) - i\beta_2\left(k_+^3\sigma_+ - \text{h.c.}\right) \\
    &+ i\beta_3\left(k_-k_+k_-\sigma_+ - \text{h.c.}\right).
  \end{split}
\end{equation}

\noindent The in-plane $g$ factor is given by

\begin{equation}\label{g_in}
\begin{split}
    \frac{g_\parallel}{2} &= \Im\left\{\Braket{c|z\textcolor{\col}{g}k_z|c} - 2\Braket{+|\textcolor{\col}{u_+^\prime}|-}\right\} \\
    &- \sqrt{2}\Re\left\{\frac{1}{\sqrt{3}\alpha_0}\Braket{c|zP|+} + \Braket{-|s}\right\},
\end{split}
\end{equation}

\noindent where

\begin{gather}
    \ket{\pm} = \ket{\ell}\pm 2^{\pm 1/2}\ket{s}, \\
    \textcolor{\col}{u_\pm^\prime} = \{z\textcolor{\col}{\gamma_3},k_z\} \textcolor{\col}{\pm} [z\textcolor{\col}{\kappa},k_z],
\end{gather}

\noindent with \textcolor{\col}{$\gamma_3$} a Luttinger parameter, \textcolor{\col}{$\kappa$} the bulk hole $g$-factor, and $P$ the so-called Kane momentum matrix element (see Appendix \ref{sec:params}). \textcolor{\col}{In the context of heterostructures, material parameters such as $\gamma_3$, $\kappa$, or $P$ are operators that act on envelope functions. Importantly, they do not commute with $k_z$ and are diagonal in position basis. For example, $\gamma_3\ket{z} = \gamma_3(z)\ket{z}$, where the function $\gamma_3(z)$ gives the value of $\gamma_3$ at coordinate $z$.} In \eqref{g_in}, the $z=0$ coordinate is chosen such that

\begin{equation}
    \braket{z} = \sum_{\tau=\{c,\ell,s\}}{\Braket{\tau|z|\tau}} = 0.
\end{equation}

\begin{figure*}[t]
    \centering
    \includegraphics[width=\linewidth]{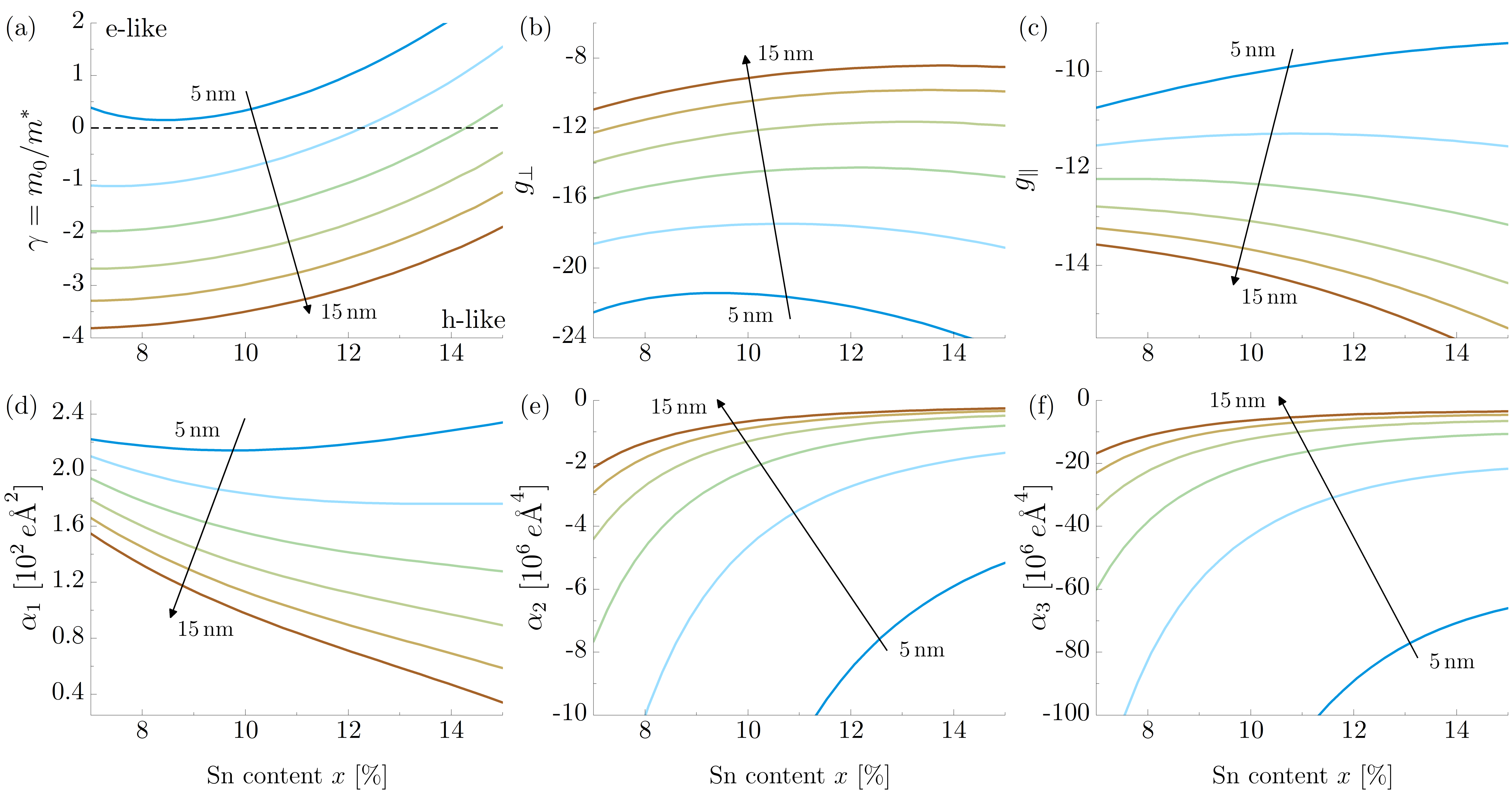}
    \caption{LH1 subband parameters as a function of the well thickness and $x$ for fully relaxed Ge$_{1-x}$Sn$_{x}$ barriers ($\varepsilon_\text{BR} = 0$). (a) Inverse effective mass $\gamma$. (b) Out-of-plane $g$ factor. (c) In-plane $g$ factor. (d), (e), and (f) are $\alpha_1$, $\alpha_2$, and $\alpha_3$ Rashba parameters, respectively. The calculations were carried out for a well thickness in the $5$--$15$ nm range. The data displayed here are for thicknesses separated by a 2 nm step.}
    \label{fig:parameters}
\end{figure*}

\noindent This ensures, when setting $k_x=k_y=0$, that $g_\parallel$ is gauge independent and corresponds to that at the subband edge. This is because the quantum numbers $k_\pm$ are generally gauge dependent, and if $\braket{z} = 0$, taking the expectation value on both sides of $K_\pm = k_\pm \mp ize^{\pm i\phi}/\lambda^2$ gives $\Braket{K_\pm} = k_\pm$, and thus associates to $k_\pm$ the gauge independent quantity $\Braket{K_\pm}$. Equation \eqref{g_in} reduces to the well-known $|g_\parallel| = 4\textcolor{\col}{\kappa}$ in the special case of 4-band Luttinger Hamiltonian with $\text{LBO}\to\infty$. The in-plane effective mass $\gamma$, the $g$-factor components, and the three Rashba parameters $\alpha_i$ are plotted as a function of $x$ and $w$ in Fig.~\ref{fig:parameters} panels (a), (b)--(c), and (d)--(f) respectively. A negative $g$ means that the spin-down level ($\sigma = -1$) is closer to the bandgap than the spin-up level.

\subsection{LH-HH mixing}

In the vicinity of $\mathbf{k}_\parallel = 0$, $\eta$ subbands acquire a small HH component in addition to the three terms in \eqref{eta}, resulting in a $\eta$-HH mixed state $\ket{\psi, \mathbf{k}_\parallel}$. For LH1, this can be written as (up to a normalization constant and to first order in $k$):

\begin{equation}\label{mix}
    \ket{\psi,\mathbf{k}_\parallel} = \ket{\eta,\sigma} - i\sigma\alpha_0 k_{-\sigma}\Ket{\frac{3}{2},\frac{3\sigma}{2}}\sum_{l}{\frac{\textcolor{\col}{T_l^\text{x}}\ket{h_l}}{E_\text{LH1} - E_l^\text{H}}} + \dots,
\end{equation}

\noindent where $\Ket{\frac{3}{2},\frac{3\sigma}{2}}$ is the HH bulk Bloch state, $\ket{h_l}$ is the $l$-th HH envelope with energy $E_l^\text{H}$ at $\mathbf{k}_\parallel = 0$, and the coefficients \textcolor{\col}{$T_l^\text{x}$} are

\begin{gather}
    \textcolor{\col}{T_l^\text{x}} = \bra{h_l}\left(\frac{P}{\sqrt{2}\alpha_0}\ket{c} - \sqrt{3}i\textcolor{\col}{u}_+\ket{-}\right),\label{Tx} \\
    \textcolor{\col}{u}_\pm = \{\textcolor{\col}{\gamma_3},k_z\} \pm [\textcolor{\col}{\kappa},k_z].
\end{gather}

To first order in $k$, mixing is stronger between $\eta$ and HH subbands with opposite parity (from $k_z$ terms in \textcolor{\col}{$T_l^\text{x}$}) and with the same spin component sign (i.e., $\ket{\eta,\sigma}$ couples with $\Ket{\frac{3}{2},\frac{3\sigma}{2}}$). The HH contribution $\rho$ in the mixed subband $\Ket{\psi,\mathbf{k}_\parallel}$ is given by the sum of the absolute square of each coefficient in front of HH terms. By symmetry, only even powers of $k$ must appear in $\rho$:

\begin{equation}\label{mix2}
    \rho = ak_\parallel^2 + O(k^4),
\end{equation}

\noindent where $a$ can be found from \eqref{mix}:

\begin{equation}
    a = \alpha_0^2\sum_l{\frac{|\textcolor{\col}{T_l^\text{x}}|^2}{\left(E_\text{LH1} - E_l^\text{H}\right)^2}}.
\end{equation}

The expression $\rho \approx ak_\parallel^2$ is valid for small $\mathbf{k}_\parallel$ such that $\rho \ll 1$. In general, $\rho$ \textcolor{\col}{lies} in the interval $[0,1]$ with $\rho = 0$ ($\rho = 1$) corresponding to a pure $\eta$ (HH) subband.

\begin{figure}[t]
    \centering
    \includegraphics[width=\linewidth]{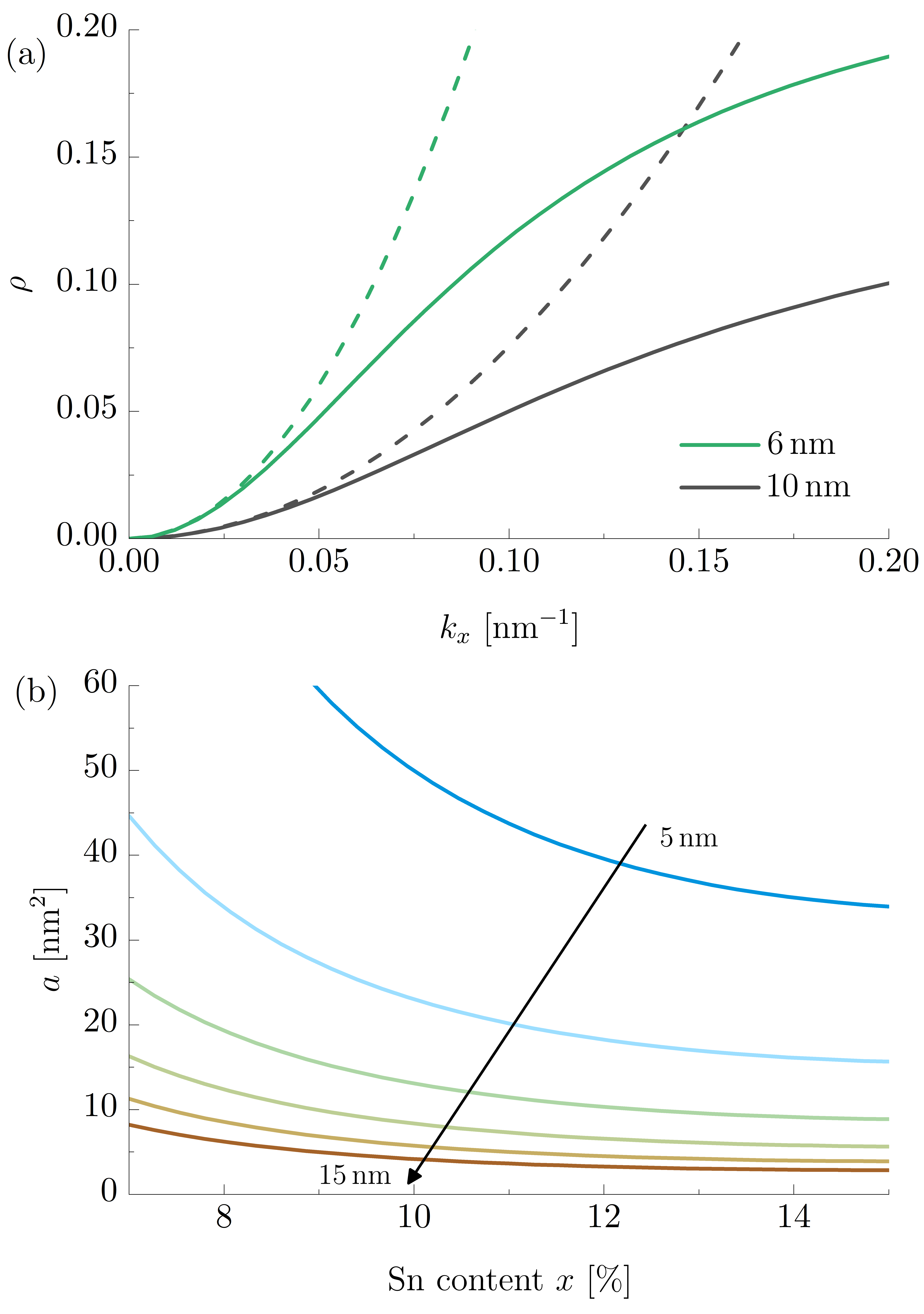}
    \caption{(a) HH contribution $\rho$ as a function of the wavevector $k_x$ for a quantum well thickness $w = 6\,\text{nm}$ and $w = 10\,\text{nm}$. Dashed lines correspond to the quadratic approximation $\rho \approx ak_\parallel^2$. (b) Coefficient $a$ as a function of $w$ and $x$. The data displayed here are for thicknesses separated by a 2 nm step.}
    \label{fig:mixing}
\end{figure}

$\rho$ \textcolor{\col}{as a function of $k_x$ is displayed in Fig.~\ref{fig:mixing}(a)} for $w=6\,\text{nm}$ and $w=10\,\text{nm}$ \textcolor{\col}{at} $x = 0.13$. When $k_x$ is small, the parabolic term in \eqref{mix2} fits well the numerically computed $\rho$. \textcolor{\col}{Mixing decreases with increasing energy splitting between LH1 and the HH continuum, as indicated by smaller $\rho$ at the larger well thickness $w=10\,\text{nm}$}. This \textcolor{\col}{remains true} for different Sn compositions, as \textcolor{\col}{illustrated} in Fig.~\ref{fig:mixing}(b) where $a$ is plotted as a function of $w$ and $x$ for $\varepsilon_\text{BR} = 0$.

\section{Discussion}
The preceding results demonstrate that the lattice mismatch between Ge$_{1-x}$Sn$_x$ alloys and Ge provides an additional degree of freedom to engineer the tensile strain required to confine LHs in Ge. According to Fig.~\ref{fig:conf}(a), the region of interest, as defined by the parameters $(x,\varepsilon_\text{BR})$, lies in the range where $x$ is above $0.12$ and the compressive strain in the barriers $\varepsilon_\text{BR}$ is below $-0.4\%$ (i.e., $|\varepsilon_\text{BR}|\lesssim 0.4\%$). In this range, all band offsets $\Delta E_i$ are positive and the Ge layer is of direct bandgap. Ge$_{1-x}$Sn$_x$ layers at Sn content in the proposed range have already been demonstrated experimentally~\cite{Moutanabbir2021}. However, the addition of a highly tensile strained Ge layer on top of strain-relaxed Ge$_{1-x}$Sn$_x$  is still under development. For instance, the authors in Ref.~\cite{Assali2022} reported a $1.65\%$ tensile-strained Ge quantum well on partially relaxed Ge$_{0.854}$Sn$_{0.146}$ barriers with a residual strain to $\varepsilon_\text{BR}\approx -0.54\%$. This system would be located near the crossing between the $0\,\text{meV}$ $\Delta E_1$ line and the $50\,\text{meV}$ LBO line in Fig.~\ref{fig:conf}(a), very close to the optimal region of interest mentioned earlier. Strain relaxation in the barriers is necessary to enhance confinement in Ge, while relaxing the criterion of minimal well thickness $w_0$ required for a LH-like valence band edge ($\Delta E_2 > 0$). The ideal amount of strain relaxation for a given barrier Sn content can be estimated from Fig.~\ref{fig:conf}(b). For example, a barrier with $x=0.14$ does not allow a LH-like valence band edge for $|\varepsilon_\text{BR}| > 0.542\%$ compressive strain. The additional requirement $w_0 < h_c$, where $h_c$ is the critical thickness of Ge, further reduces the range of $|\varepsilon_\text{BR}|$ to around $< 0.5\%$ compressive strain. Reducing the amount of Sn in the barriers relaxes the limit imposed by the critical thickness $h_c$, but at the cost of a smaller LBO. In contrast, increasing $x$ to $0.18$ for instance slightly increases the range for $|\varepsilon_\text{BR}|$ to around $<0.525\%$, and increases significantly the LBO and $\Delta E_1$ [Fig.~\ref{fig:conf}(a)] but at the expense of a smaller $h_c$ and a narrower window $w_0 < w < h_c$, but still in the range typically achievable in epitaxial growth experiments.

The in-plane effective mass $\gamma$ [Fig. \ref{fig:parameters}(a)] shows a strong dependence on both $x$ and $w$, with small \textcolor{\col}{$\gamma$} expected from the general rule that LHs are heavier in the plane than HHs. There is also an interesting feature where the dispersion changes from a hole-like ($\gamma < 0$) to an electron-like ($\gamma > 0$) curvature at $\mathbf{k}_\parallel=0$. In the hole-like regime, the valence band edge is formed by a single valley located at $\mathbf{k}_\parallel = 0$. In contrast, in the electron-like regime, the valence band edge consists of four valleys, each located a distance $k_0^*$ from $\mathbf{k}_\parallel = 0$ along the four equivalent $\left<110\right>$ crystallographic directions in the QW plane. For instance, for a $5\,\text{nm}$ well at $x = 0.13$ and $\varepsilon_\text{BR} = 0$, the valley minima are located at $k_0^* \approx 0.0813\,\text{nm}^{-1}$ away from $\mathbf{k}_\parallel = 0$. For larger wavevectors, the dispersion goes away from the bandgap as required, owing to $k^4$-terms or higher that are not taken in account by the effective Hamiltonians \eqref{eff_par} and \eqref{eff_in}. According to Fig. \ref{fig:parameters}(a), electron-like subbands occur for small $w$ and large $x$. This effect takes place for two reasons. The first is when the LH subband anticrosses a neighboring HH subband such that the curvature is inverted at $\mathbf{k}_\parallel = 0$. This is typical in systems where the ground state is HH-like and the first LH subband is allowed to mix strongly with the first excited HH subband~\cite{Hayden1991}, or when the LH subband is close to a HH continuum (e.g., when $w$ is small). When a LH is far from neighboring HH levels (e.g., when tensile strain is large), mixing decreases, as illustrated in Fig.~\ref{fig:mixing}(b), and becomes too weak to invert the curvature. The second reason for a curvature change is related to the anticrossing between the LH and the SO bands in the bulk dispersion of Ge for $k_z > 0$ [Fig.~\ref{fig:stack}(a)]. This anticrossing results in a curvature sign change of the LH band at some point $k_z^*$ such that the bulk energy dispersion $E(k_x,k_y;k_z)$ has a hole-like (electron-like) curvature when $k_z$ is fixed to a value smaller (larger) than $k_z^*$ and $k_{x,y}$ are close to zero. For a quantum well along the $z$ direction, the reciprocal-space envelope functions $\tilde{\psi}(k_z)$ become wider for thinner wells and thus get a larger contribution from the electron-like regions in $k$-space, resulting in a dispersion with inverted curvature at $\mathbf{k}_\parallel = 0$.

A major difference with $\eta$ subbands is the large $g_\parallel$ compared to HH systems~\cite{Jirovec2022,Watzinger2018,Hendrickx2018,Hendrickx2020_2,Lawrie2020}. \textcolor{\col}{A comparison between $g_\parallel$ and $g_\perp$ reveals an anisotropy} for well thicknesses away from ${\sim} 10\,\text{nm}$. Both components have a stronger dependence on $w$ than on $x$ but have opposite behavior with $w$ due to how they couple with neighboring subbands. For the out-of-plane component, $g_\perp\sim 2\textcolor{\col}{\kappa}$ for large $w$ because the coupling with neighboring levels becomes weaker as LH1 \textcolor{\col}{gets further} from the continuum. In contrast, the in-plane $g$-factor does not depend on couplings with neighboring HH levels [c.f. \eqref{g_in}] and is instead more influenced by the spatial distribution of the envelopes across the layers. Thus, for large (small) $w$, $g_\parallel\sim 4\textcolor{\col}{\kappa}$ with \textcolor{\col}{$\kappa$ being that of} Ge (Ge$_{1-x}$Sn$_{x}$).

Another peculiar feature associated with $\eta$ subbands is the absence of direct connection between $\gamma$ and $g_\perp$ \textcolor{\col}{in contrast with} HHs (i.e., see Eq. (5) in Ref.~\cite{Drichko2018}). \textcolor{\col}{Perturbation theory gives the following for} $\gamma$ and $g_\perp$:

 \begin{subequations}\label{pert_eta}
 \begin{align}
     \gamma &= \textcolor{\col}{\Gamma^\eta} + \textcolor{\col}{C} + \textcolor{\col}{D}, \\
     \frac{g_\perp}{2} &= \frac{\textcolor{\col}{G^\eta}}{2} - \textcolor{\col}{C} + \textcolor{\col}{D},
 \end{align}
\end{subequations}

\noindent where \textcolor{\col}{$\Gamma^\eta$, $G^\eta$ are described in Appendix~\ref{sec:model}, \textcolor{\col}{whereas $C$ and $D$} are described in Appendix~\ref{sec:pert}}. For HHs, a similar expansion would give

 \begin{subequations}\label{pert_HH}
 \begin{align}
     \gamma^\text{H} &= \textcolor{\col}{\Gamma^\text{H}} + \textcolor{\col}{C^\prime}, \\
     \frac{g_\perp^\text{H}}{2} &= \frac{\textcolor{\col}{G^\text{H}}}{2} + \textcolor{\col}{C^\prime}.
 \end{align}
\end{subequations}

\noindent In the latter case, one can combine the equations for $\gamma^\text{H}$ and $g_\perp^\text{H}$ to eliminate the \textcolor{\col}{$C^\prime$} term, resulting in an expression involving only the mass and the $g$-factor~\cite{Drichko2018,Wimbauer1994}:

\begin{equation}
    \frac{g_\perp^\text{H}}{2} = \frac{\textcolor{\col}{G^\text{H}}}{2} - \textcolor{\col}{\Gamma^\text{H}} + \gamma^\text{H}.
\end{equation}

However, the result is different for $\eta$ subbands due to the additional \textcolor{\col}{$D$} term in \eqref{pert_eta}. The latter is also related to the nonzero $\beta_1$ coefficient of $\eta$ subbands~\cite{DelVecchio2023}.

Rashba parameters \textcolor{\col}{follow} the general behavior $\alpha_i\to 0$ as $w$ increases. \textcolor{\col}{This is} caused by \textcolor{\col}{a} reduced \textcolor{\col}{sensitivity of the wavefunction to electric fields when the level does not spread as extensively into the barriers.} Although the \textcolor{\col}{QW} is characterized by relatively small LBOs ($\lesssim 100\,\text{meV}$) and small out-of-plane \textcolor{\col}{LH effective masses}, the device operation can comfortably sustain realistic DC electric fields along the growth direction regardless of the well thickness and LBO without inducing any envelope leak into the barriers. In devices where space inversion symmetry needs to be broken, such as in electric dipole spin resonance (EDSR) experiments, this should not be an issue as the relevant Rashba parameter $\alpha_3$ for EDSR, (which is proportional to $\textcolor{\col}{\gamma_2} + \textcolor{\col}{\gamma_3}$) is one order of magnitude larger than the $\alpha_2$ Rashba parameter (proportional to $\textcolor{\col}{\gamma_2} - \textcolor{\col}{\gamma_3}$), thus requiring smaller out-of-plane fields~\cite{DelVecchio2023}.

\section{Conclusions}

This work \textcolor{\col}{demonstrates} how Ge$_{1-x}$Sn$_x$/Ge/Ge$_{1-x}$Sn$_x$ heterostructures can be tailored to achieve \textcolor{\col}{a} selective confinement of LHs in Ge \textcolor{\col}{while pushing HHs in to the Ge$_{1-x}$Sn$_x$ barriers}. For a sufficiently large Sn content ($x>0.12$), small residual compressive strain in the barriers ($|\varepsilon_\text{BR}|<0.4\%$) \textcolor{\col}{and a well thickness $w > w_0$}, the LH ground state \textcolor{\col}{emerges from within the HH continuum}, thus yielding a pure LH-like valence band edge ($\Delta E_2 > 0$). This regime also corresponds to a direct bandgap in both the well and its barriers, owing to the large tensile strain in the Ge well and the high Sn content in the barriers. Satisfying the condition of $\Delta E_2 > 0$ imposes a threshold for residual strain in the barriers beyond which a LH-like VB edge becomes virtually impossible.

The in-plane effective mass, the out-of-plane and in-plane $g$-factor, and the Rashba parameters $\alpha_{1,2,3}$ were computed by explicitly taking in account the spread of the LH envelopes into the barriers and the coupling with the neighboring HH continuum. \textcolor{\col}{Small inverse effective masses $\gamma$ are obtained}. A peculiar sign change in $\gamma$ appearing for small well thicknesses ($w\lesssim 7\,\text{nm}$) is observed and attributed to the proximity of the LH to the HH continuum (larger LH-HH mixing) and the contribution of the SO band in the LH spinor. An increasingly strong anisotropy in the $g$-factor components is also observed for well thicknesses away from ${\sim} 10\,\text{nm}$. Most notably, the in-plane component of the $g$-tensor is significantly larger than what is expected in HH systems. \textcolor{\col}{A nonzero linear Rashba parameter $\alpha_1$ was obtained, as anticipated for LH systems}, with an $\alpha_3$ coefficient one order of magnitude larger than $\alpha_2$.

\medskip
\noindent {\textbf{Acknowledgments}}.
O.M. acknowledges support from NSERC Canada (Discovery, SPG, and CRD Grants), Canada Research Chairs, Canada Foundation for Innovation, Mitacs, PRIMA Qu\'ebec, Defence Canada (Innovation for Defence Excellence and Security, IDEaS), the European Union’s Horizon Europe research and innovation program under Grant Agreement No 101070700 (MIRAQLS), the US Army Research Office Grant No. W911NF-22-1-0277, and the Air Force Office of Scientific and Research Grant No. FA9550-23-1-0763. \\

\appendix
\section{Parameterization of G\lowercase{e}\textcolor{\col}{$_{1-x}$}S\lowercase{n}\textcolor{\col}{$_x$}}\label{sec:params}

The material parameters for the Ge$_{1-x}$Sn$_x$ alloy were calculated in the full composition range by interpolating the parameters from pure Ge and pure Sn:

\begin{equation}\label{Vegard}
  A(x) = (1-x)A^\text{Ge} + xA^\text{Sn} - x(1-x)b.
\end{equation}

\begin{table}[t]
\begin{threeparttable}
    \caption{Input parameters with bowings for the 8-band $k\cdot p$ model.}\label{tb:parameters}
    \begin{tabular}{l c c c}
        \toprule
        Parameter & Germanium & Tin & Bowing \\
        \midrule
        Lattice constant & & & \\
        \hline
        $a_0$ (\AA, $300\,\text{K}$) & $5.657956$\tnote{a} & $6.489417$\tnote{b} & \ $-0.083$\tnote{c*} \\
        \hline
        Energy gaps & & & \\
        \hline
        $E_{g\Gamma}^0$ (eV) & $0.8981$\tnote{b} & $-0.413$\tnote{b} & $2.46$\tnote{d*} \\
        $E_{g\text{L}}^0$ (eV) & $0.740$\tnote{p} & $0.100$\tnote{c} & $1.23$\tnote{k*}\\
        $\alpha_\Gamma$ ($10^{-4}\,\text{eV}/\,\text{K}$) & $6.842$\tnote{e} & $-7.94$\tnote{d} & \\
        $\alpha_\text{L}$ ($10^{-4}\,\text{eV}/\,\text{K}$) & $4.561$\tnote{e} & & \\
        $\beta_\Gamma$ (K) & $398$\tnote{e} & $11$\tnote{d} & \\
        $\beta_\text{L}$ (K) & $210$\tnote{e} & & \\
        $\Delta$ (eV) & $0.290$\tnote{b} & $0.770$\tnote{f} & $-0.100$\tnote{c} \\
        $E_{v,\text{avg}}$ (eV) & $0$ & $0.69$\tnote{g} & \\
        \hline
        Elastic constants & & & \\
        \hline
        $c_{11}$ (GPa) & $124$\tnote{b} & $69.0$\tnote{b} & \\
        $c_{12}$ (GPa) & $41.3$\tnote{b} & $29.3$\tnote{b} & \\
        $c_{44}$ (GPa) & $68.3$\tnote{b} & $36.2$\tnote{b} & \\
        \hline
        \multicolumn{4}{l}{Deformation potentials\tnote{$\dag$}} \\
        \hline
        $a_{c\Gamma}$ (eV) & $-8.24$\tnote{h} & $-6.00$\tnote{l} & \\
        $a_{c\text{L}}$ (eV) & $-1.54$\tnote{h} & $-2.14$\tnote{m} & \\
        $a_v$ (eV) & $1.24$\tnote{h} & $1.58$\tnote{m} & \\
        $b$ (eV) & $-2.86$\tnote{i} & $-2.7$\tnote{n} & \\
        \hline
        \multicolumn{4}{l}{Effective mass and spin parameters} \\
        \hline
        $m_{c\Gamma}^\text{\textcolor{\col}{L}}$ ($m_0$) & $0.0386$ & $-0.057$ & \\
        $\gamma_1^\text{\textcolor{\col}{L}}$ & $13.38$\tnote{j} & & \\
        $\gamma_2^\text{\textcolor{\col}{L}}$ & $4.24$\tnote{j} & \multicolumn{2}{c}{See Eq. \eqref{gamma}} \\
        $\gamma_3^\text{\textcolor{\col}{L}}$ & $5.69$\tnote{j} & & \\
        $\kappa^\text{\textcolor{\col}{L}}$ & $3.41$\tnote{f} & $-11.84$\tnote{f} & \\
        $g^\text{\textcolor{\col}{L}}$ & $-2.77$ & $86.6$ & \\
        \bottomrule
    \end{tabular}
    \begin{tablenotes}
    \item [$\dag$] The convention $a = a_c - a_v$ is used.
    \item [*] See Eq. \eqref{bandgap}
    \item [a] Reference \cite{Reeber1996}
    \item [b] Reference \cite{Madelung1991}
    \item [c] Reference \cite{Polak2017}
    \item [d] Reference \cite{Bertrand2019}
    \item [e] Reference \cite{Varshni1967}
    \item [f] Reference \cite{Lawaetz1971}
    \item [g] Reference \cite{Menendez2004}
    \item [h] Reference \cite{VandeWalle1989}
    \item [i] Reference \cite{VandeWalle1986}
    \item [j] Reference \cite{Winkler2003}
    \item [k] Reference \cite{DCosta2006}
    \item [l] Reference \cite{Li2006}
    \item [m] Reference \cite{Brudevoll1993}
    \item [n] Reference \cite{Willatzen1995}
    \item [p] Reference \cite{Weber1989}
    \end{tablenotes}
    \end{threeparttable}
\end{table}

Here, $0 \leq x \leq 1$ is the alloy fraction and $b$ is a bowing constant for the parameter $A$ if necessary. For temperature-dependent quantities such as the lattice constant or the bandgaps, we apply \eqref{Vegard} on the temperature-dependent parameters of the alloy's constituents. For instance, the bandgap of Ge$_{1-x}$Sn$_x$ at the $\Gamma$ point, $E_{g\Gamma}(x,T)$, is given by 

\begin{equation}\label{bandgap}
    E_{g\Gamma}(x,T) = (1-x)E_{g\Gamma}^\text{Ge}(T) + xE_{g\Gamma}^\text{Sn}(T) - x(1-x)b_\Gamma,
\end{equation}

\noindent where

\begin{equation}
    E_{g\Gamma}^\text{Ge}(T) = E_{g\Gamma}^{0,\text{Ge}} - \frac{\alpha_\Gamma^\text{Ge}T^2}{\beta_\Gamma^\text{Ge} + T}
\end{equation}

\noindent and similar equations for Sn and the L valley. The average VB energy $E_{v,\text{avg}}$ is given by $E_{v,\text{avg}}=E_v-\Delta/3$, where $E_v$ is the VB edge energy and $\Delta$ is the spin-orbit splitting~\cite{VandeWalle1989}.

The electron effective mass $m_{c\Gamma}^\text{\textcolor{\col}{L}}$ and the electron $g^\text{\textcolor{\col}{L}}$ factor were calculated following the approach in Ref. \cite{Lawaetz1971} to make them consistent with the bandgaps and spin-orbit couplings listed in Table~\ref{tb:parameters}. The Luttinger parameters $\gamma_{1,2,3}^\text{\textcolor{\col}{L}}$ were interpolated between pure Ge and Ge$_{0.80}$Sn$_{0.20}$ using the data from Ref. \cite{LuLow2012}, giving

\begin{equation}\label{gamma}
    \gamma_i^\text{\textcolor{\col}{L}} = \gamma_i^\text{\textcolor{\col}{L},Ge}\left(1 - \frac{x}{0.2}\right) + \gamma_i^\text{\textcolor{\col}{L},GeSn}\left(\frac{x}{0.2}\right) - b_i\frac{x}{0.2}\left(1-\frac{x}{0.2}\right),
\end{equation}

with $\gamma_i^\text{\textcolor{\col}{L},Ge}$ listed in Table \ref{tb:parameters}, $\gamma_1^\text{\textcolor{\col}{L},GeSn} = 29.2108$, $\gamma_2^\text{\textcolor{\col}{L},GeSn} = 12.2413$, $\gamma_3^\text{\textcolor{\col}{L},GeSn} = 13.7387$, and $b_1 = 20.3391$, $b_2 = 9.6609$, $b_3 = 9.8187$.

The Kane momentum matrix element $P$, which couples the two conduction bands to the six valence bands, is known to sometimes cause spurious solutions, often appearing as levels within the bandgap or with violently oscillating envelopes~\cite{Foreman1997,Veprek2007,Eissfeller2011}. The approach employed here to eliminate spurious solutions is described in Ref.~\cite{Foreman1997} and consists of rescaling $P$:

\begin{equation}\label{rescale}
    P^2 = \frac{3\hbar^2}{2m_{c\Gamma}^\text{\textcolor{\col}{L}}}\left(\frac{2}{E_{g\Gamma}} + \frac{1}{E_{g\Gamma} + \Delta}\right)^{-1}.
\end{equation}

The remote band contributions in the Luttinger parameters are then re-ajusted~\cite{Winkler2003} according to $P \equiv \sqrt{\alpha_0 E_p}$ given in \eqref{rescale}:

\begin{gather}
  \textcolor{\col}{\gamma_1} = \gamma_1^\text{\textcolor{\col}{L}} - \frac{E_p}{3E_g} \\
  \textcolor{\col}{\gamma_{2,3}} = \gamma_{2,3}^\text{\textcolor{\col}{L}} - \frac{E_p}{6E_g} \\
  \textcolor{\col}{\kappa} = \kappa^\text{\textcolor{\col}{L}} - \frac{E_p}{6E_g} \\
  \textcolor{\col}{g} = g^\text{\textcolor{\col}{L}} + \frac{2E_p}{3E_g}\frac{\Delta}{E_g + \Delta} \\
  \frac{m_0}{\textcolor{\col}{m_{c\Gamma}}} = \frac{m_0}{m_{c\Gamma}^\text{\textcolor{\col}{L}}} - \frac{2E_p}{3E_g}\frac{3E_g/2 + \Delta}{E_g + \Delta} = 0.\label{A0}
\end{gather}

\color{\col}
\section{$k\cdot p$ framework}\label{sec:model}

Our implementation of $k\cdot p$ theory is based on the model presented in Ref. \cite{Eissfeller2012}, which is an extension of standard $k\cdot p$ frameworks~\cite{Winkler2003,Voon2009} for heterostructures with finite energy band offsets at the interfaces (i.e., when a proper ordering between material parameter operators and $k_z$ is critical).

The 8-band $k\cdot p$ Hamiltonian can be written as a sum of different contributions:

\begin{equation}\label{H}
    H = H_k + H_\text{SO} + H_\varepsilon + V.
\end{equation}

The first term, $H_k$, depends on the mechanical wavevector $\textbf{K} = \textbf{k} + e\textbf{A}/\hbar$. It automatically includes the Zeeman Hamiltonian through the relation $\textbf{K}\times \textbf{K} = e\textbf{B}/(i\hbar)$. In the Cartesian basis

\begin{equation}
\begin{split}
    \mathcal{B}_X = &\left\{\ket{S+},\ket{S-},\right. \\
    &\left.\ket{X+},\ket{Y+},\ket{Z+},\ket{X-},\ket{Y-},\ket{Z-}\right\},
\end{split}
\end{equation}

\noindent $H_k$ is given by

\begin{equation}
\begin{split}
    H_{k} &=\begin{bmatrix}
    H_{cc}^{k} & 1_{2\times 2} \otimes H_{cv}^{k}\\
    1_{2\times 2} \otimes H_{cv}^{k\dagger} & 1_{2\times 2} \otimes H_{vv}^{k}
  \end{bmatrix} \\
  &+\begin{bmatrix}
    H_{B} & 0\\
    0 & H_{B} \otimes 1_{3\times 3}
  \end{bmatrix},
\end{split}
\end{equation}

\noindent with

\begin{gather}
\begin{split}
    H_{cc}^{k} &= E_{g} +\sum _{\alpha } K_{\alpha } AK_{\alpha } \\
    &+\frac{i\alpha _{0}}{2}\sum _{\alpha \beta \gamma } \epsilon _{\alpha \beta \gamma } K_{\alpha }( g-g_{0}) K_{\beta } \sigma _{\gamma },
\end{split} \\
  H_{cv}^{k} =\begin{bmatrix}
    iPK_{x} & iPK_{y} & iPK_{z}
  \end{bmatrix}, \\
  \left( H_{vv}^{k}\right)_{i,j} =\begin{cases}
    \sum _{\alpha } K_{\alpha } MK_{\alpha } +K_{i}( L-M) K_{i} & i=j\\
    K_{i} N_{+} K_{j} +K_{j} N_{-} K_{i} & i\neq j
  \end{cases}, \\
  H_{B} =\frac{i\alpha _{0} g_{0}}{2}\sum _{\alpha \beta \gamma } \epsilon _{\alpha \beta \gamma } K_{\alpha } K_{\beta } \sigma _{\gamma },
\end{gather}

\noindent where $\epsilon_{\alpha\beta\gamma}$ is the Levi-Civita tensor, $A = \alpha_0m_0/m_{c\Gamma} = 0$ [see \eqref{A0}] and $g_0\approx 2$ is the free electron $g$-factor. The term $H_B$ arises from the interaction of the free electron spin with the magnetic field~\cite{Eissfeller2012}.

The second term in~\eqref{H}, $H_\text{SO}$, includes the effects of the SO band and is proportional to the spin-orbit gap $\Delta$:

\begin{equation}
    H_\text{SO} = \frac{\Delta }{3}\begin{bmatrix}
1 & 0 & 0 & 0 & 0 & 0 & 0 & 0\\
0 & 1 & 0 & 0 & 0 & 0 & 0 & 0\\
0 & 0 & 0 & -i & 0 & 0 & 0 & 1\\
0 & 0 & i & 0 & 0 & 0 & 0 & -i\\
0 & 0 & 0 & 0 & 0 & -1 & i & 0\\
0 & 0 & 0 & 0 & -1 & 0 & i & 0\\
0 & 0 & 0 & 0 & -i & -i & 0 & 0\\
0 & 0 & 1 & i & 0 & 0 & 0 & 0
\end{bmatrix}.
\end{equation}

Strain is incorporated by means of the Bir-Pikus formalism~\cite{Bir1974}, resulting in the term

\begin{equation}
H_{\varepsilon } =\begin{bmatrix}
H_{cc}^{\varepsilon } & 0\\
0 & 1_{2\times 2} \otimes H_{vv}^{\varepsilon }
\end{bmatrix},
\end{equation}

\noindent with

\begin{gather}
H_{cc}^{\varepsilon } =1_{2\times 2} a_{c}\text{Tr}\{\varepsilon \}, \\
\left( H_{vv}^{\varepsilon }\right)_{i,j} =\begin{cases}
m\text{Tr}\{\varepsilon \} +( l-m) \varepsilon _{ii} & i=j\\
n\varepsilon _{ij} & i\neq j
\end{cases}.
\end{gather}

Finally, the potential energy $V = E_{v,\text{avg}} + eE_z z$. The parameters $L$, $M$, $N_\pm$, $l$, $m$, and $n$ are related to the usual Luttinger parameters and deformation potentials by

\begin{gather}
\begin{bmatrix}
L\\
M\\
N_{+} +\alpha _{0}\\
N_{-} -\alpha _{0}
\end{bmatrix} =-\alpha _{0}\begin{bmatrix}
1 & 4 & 0 & 0\\
1 & -2 & 0 & 0\\
0 & 0 & 3 & 3\\
0 & 0 & 3 & -3
\end{bmatrix}\begin{bmatrix}
\gamma _{1}\\
\gamma _{2}\\
\gamma _{3}\\
\kappa 
\end{bmatrix}, \\
\begin{bmatrix}
l\\
m\\
n
\end{bmatrix} =\begin{bmatrix}
1 & 2 & 0\\
1 & -1 & 0\\
0 & 0 & \sqrt{3}
\end{bmatrix}\begin{bmatrix}
a_{v}\\
b\\
d
\end{bmatrix}.
\end{gather}

A change of basis from $\mathcal{B}_X$ to the so-called angular momentum basis $\mathcal{B}_J$ brings the Hamiltonian $H$ in a $2\times 2$ block diagonal matrix (each block being $4\times 4$) when evaluated with $B=0$ and $k_x = k_y = 0$:

\begin{align*}
    H_0 &\equiv U_{X\leftarrow J}^\dagger H(k_x=k_y=0, B=0)U_{X\leftarrow J} \\
    &= \begin{bmatrix}
        H_+ & 0 \\
        0 & H_-
    \end{bmatrix},
\end{align*}
    
\noindent where

\begin{gather}
\begin{split}
    \mathcal{B}_J = &\left\{\Ket{\frac{1}{2},\frac{1}{2}}_c, \Ket{\frac{3}{2},\frac{1}{2}}, \Ket{\frac{1}{2},\frac{1}{2}}, \Ket{\frac{3}{2},\frac{3}{2}},\right. \\
    &\left.\Ket{\frac{1}{2},-\frac{1}{2}}_c, \Ket{\frac{3}{2},-\frac{1}{2}}, \Ket{\frac{1}{2},-\frac{1}{2}}, \Ket{\frac{3}{2},-\frac{3}{2}}\right\},
\end{split} \\
    U_{X\leftarrow J} =\begin{bmatrix}
1 & 0 & 0 & 0 & 0 & 0 & 0 & 0\\
0 & 0 & 0 & 0 & 1 & 0 & 0 & 0\\
0 & 0 & 0 & -s_{2} & 0 & s_{6} & -s_{3} & 0\\
0 & 0 & 0 & -is_{2} & 0 & -is_{6} & is_{3} & 0\\
0 & s_{23} & -s_{3} & 0 & 0 & 0 & 0 & 0\\
0 & -s_{6} & -s_{3} & 0 & 0 & 0 & 0 & s_{2}\\
0 & -is_{6} & -is_{3} & 0 & 0 & 0 & 0 & -is_{2}\\
0 & 0 & 0 & 0 & 0 & s_{23} & s_{3} & 0
\end{bmatrix}, \\
\begin{split}
    H_{\sigma } &=V +\frac{\Delta }{3} + \begin{bmatrix}
        E_{g\Gamma} & 0 & 0 & 0 \\
        & 0 & 0 & 0 \\
        \dagger & & -\Delta & 0 \\
        & & & 0
    \end{bmatrix} \\
    &+\begin{bmatrix}
k_{z} Ak_{z} & i\sqrt{2/3} Pk_{z} & -i\sigma Pk_{z} /\sqrt{3} & 0\\
 & -\alpha _{0} k_{z} \gamma _{+} k_{z} & \sqrt{8} \sigma \alpha _{0} k_{z} \gamma _{2} k_{z} & 0\\
\dagger  &  & -\alpha _{0} k_{z} \gamma _{1} k_{z} & 0\\
 &  &  & -\alpha _{0} k_{z} \gamma _{-} k_{z}
\end{bmatrix} \\
&+ \text{Tr}\{\varepsilon \}\begin{bmatrix}
a_{c} & 0 & 0 & 0\\
 & a_{v} & 0 & 0\\
\dagger &  & a_{v} & 0\\
 &  &  & a_{v}
\end{bmatrix}
+b\delta \varepsilon \begin{bmatrix}
0 & 0 & 0 & 0\\
 & -1 & \sqrt{2} \sigma  & 0\\
\dagger & & 0 & 0\\
 & & & 1
\end{bmatrix},
\end{split}
\end{gather}

\noindent with $s_2 = 1/\sqrt{2}$, $s_3 = 1/\sqrt{3}$, $s_6 = 1/\sqrt{6}$, $s_{23} = \sqrt{2/3}$, $\sigma = \pm 1$ the pseudo-spin index, $\gamma_\pm = \gamma_1\pm 2\gamma_2$, and $\delta\varepsilon = \varepsilon_{xx} - \varepsilon_{zz}$ (all shear components in the strain tensor vanish). Since $H_+$ and $H_-$ differ only by a minus sign in the LH-SO and the CB-SO coupling elements, both share the same energy spectrum (Kramers' degeneracy) and the eigenstates of $H_-$ are the time-reversed eigenstates of $H_+$. Additionally, $H_\sigma$ is itself block diagonal: one $3\times 3$ block representing a CB-LH-SO superposition (or a $\eta$ level) and one $1\times 1$ block representing a pure HH level (or H for short). Eigenstates of $H_\sigma$ are thus either of type $\eta$ or H (with pseudo-spin $\sigma$):

\begin{subequations}\label{edges}
    \begin{gather}
    \Ket{\eta,\sigma} = \Ket{\frac{1}{2},\frac{\sigma}{2}}_c\Ket{c} + \Ket{\frac{3}{2},\frac{\sigma}{2}}\Ket{\ell} + \sigma\Ket{\frac{1}{2},\frac{\sigma}{2}}\Ket{s}, \\
    \Ket{\text{H},\sigma} = \Ket{\frac{3}{2},\frac{3\sigma}{2}}\Ket{h}.
\end{gather}
\end{subequations}

The energies and eigenstates of $H_+$ are computed for each set of quantum well parameters ($x$, $\varepsilon_\text{BR}$ and $w$) with the substitution $k_z\to -i\partial_z$, without any assumptions on the shape of the envelopes. We choose a grid spacing of $0.01\,\text{nm}$ for the finite differences and keep the $200$ subbands that are the closest to LH1 ($N = N_\eta$ + $N_\text{H} = 200$). We found that $N=200$ is large enough for the effective parameters in \eqref{eff_par} and \eqref{eff_in} to converge.

We diagonalize $H$ away from $k_x = k_y = 0$ and $B = 0$ by first projecting $H$ onto the orthonormal basis \eqref{edges}. This brings $H$ to a $4\times 4$ block-matrix form (with total dimension $2N\times 2N$), where each block consists of all the subbands of one aforementioned type (H or $\eta$, spin up/down) and couplings thereof. Taking the basis ordering $\mathcal{B}_0 = \{\ket{\text{H}+},\ket{\eta +},\ket{\eta -},\ket{\text{H}-}\}$ (and bold characters to emphasize that we are in basis $\mathcal{B}_0$), the Hamiltonian $H$ when $\textbf{B}$ is perpendicular to the plane is given by

\begin{equation}
\begin{split}
    \textbf{H} = \mathbf{E}_{0} &+ \alpha _{0}\left[\mathbf{M}_{\gamma } K_{\parallel }^{2} +\frac{1}{2\lambda^{2}}\mathbf{M}_{g}\right. \\
    &+ \left.\left( i\mathbf{M}_{1} K_{-} +\mathbf{M}_{2} K_{-}^{2} +\text{h.c.}\right)\right],
\end{split}
\end{equation}

\noindent where $\textbf{E}_0 = \text{diag}\{\textbf{E}^\text{H},\textbf{E}^\eta,\textbf{E}^\eta,\textbf{E}^\text{H}\}$ are the energies associated to $H_0$ ($\textbf{E}^{\text{H},\eta}$ are also diagonal) and

\begin{subequations}\label{Mi}
    \begin{align}
    \mathbf{M}_{\gamma } &= \text{diag}\{\boldsymbol{\Gamma}^{\text{H}},\boldsymbol{\Gamma }^{\eta },\boldsymbol{\Gamma }^{\eta },\boldsymbol{\Gamma }^{\text{H}}\}, \\
\mathbf{M}_{g} &=\text{diag}\{\mathbf{G}^{\text{H}},\mathbf{G}^{\eta },-\mathbf{G}^{\eta },-\mathbf{G}^{\text{H}}\}, \\
\mathbf{M}_{1} &=\begin{bmatrix}
0 & \mathbf{T}^{\text{x}} & 0 & 0\\
0 & 0 & \mathbf{T}^{\eta } & 0\\
0 & 0 & 0 & \mathbf{T}^{\text{x}\dagger }\\
0 & 0 & 0 & 0
\end{bmatrix}, \\
\mathbf{M}_{2} &=\begin{bmatrix}
0 & 0 & \boldsymbol{\upmu } & 0\\
0 & 0 & 0 & \boldsymbol{\upmu }^{\dagger }\\
\boldsymbol{\updelta }^{\dagger } & 0 & 0 & 0\\
0 & \boldsymbol{\updelta } & 0 & 0
\end{bmatrix},
\end{align}
\end{subequations}

\noindent with (assuming $A = 0$ and $g_0= 2$):

\begin{subequations}
    \begin{gather}
\Gamma_{l,l^\prime}^{\text{H}} =-\bra{h_l} \gamma _{1} +\gamma _{2}\ket{h_{l^\prime}},\\
G_{l,l^\prime}^{\text{H}} =-\bra{h_l} 6\kappa\ket{h_{l^\prime}},\\
\Gamma_{j,j^\prime}^{\eta} =-\frac{1}{3}\bra{+_j} \gamma _{1} +\gamma _{2}\ket{+_{j^\prime}} -\frac{2}{3}\bra{-_j} \gamma _{1} -2\gamma _{2}\ket{-_{j^\prime}},\\
\begin{split}
    G_{j,j^\prime}^{\eta} =&\bra{c_j} g\ket{c_{j^\prime}} -2\bra{+_j} \kappa \ket{+_{j^\prime}} \\
    &-\frac{4}{3}\left(\braket{+_j|+_{j^\prime}} -\braket{-_j|-_{j^\prime}}\right),
\end{split} \\
\begin{split}
    T_{j,j^\prime}^{\eta} =&\frac{1}{\sqrt{6} \alpha _{0}}\left(\bra{c_j} P\ket{+_{j^\prime}} +\bra{+_j} P\ket{c_{j^\prime}}\right) \\
    &+i\bra{c_j}[g/2,k_{z}]\ket{c_{j^\prime}} \\
    &-i\left(\bra{+_j} u_{+}\ket{-_{j^\prime}} - \bra{-_j} u_{-}\ket{+_{j^\prime}}\right),
\end{split} \\
T_{l,j}^{\text{x}} =\bra{h_l}\left(\frac{P}{\sqrt{2} \alpha _{0}}\ket{c_j} -\sqrt{3} iu_{+}\ket{-_j}\right), \\
\mu_{l,j} =\frac{\sqrt{3}}{2}\bra{h_l} \gamma_2+\gamma_3 \ket{+_j}, \\
\delta_{l,j} =\frac{\sqrt{3}}{2}\bra{h_l} \gamma_2-\gamma_3 \ket{+_j}.
\end{gather}
\end{subequations}

Here, $T_{l,j}^\text{x}$ really is the same as in \eqref{Tx} but with the explicit dependence on the $\eta$ subband index $j$. When $\textbf{B}$ is in-plane, the Hamiltonian $H$ is given by

\begin{equation}\label{H_in}
\begin{split}
    \mathbf{H} = \mathbf{E}_{0} &+\alpha _{0}\left\{\mathbf{M}_{\gamma } k_{\| }^{2} +\frac{1}{\lambda ^{4}}\mathbf{M}_{\gamma }^{\prime \prime }\right. \\
    &+ \left.\left[ i\left(\mathbf{M}_{1} +\frac{2}{\lambda ^{2}}\mathbf{M}_{2}^{\prime } e^{-i\phi } -\frac{1}{\lambda ^{2}}\mathbf{M}_{\gamma }^{\prime } e^{i\phi }\right) k_{-}\right.\right. \\
    &+\left.\left.\mathbf{M}_{2} k_{-}^{2} -\frac{1}{\lambda ^{2}}\mathbf{M}_{1}^{\prime } e^{-i\phi } -\frac{e^{-2i\phi }}{\lambda ^{4}}\mathbf{M}_{2}^{\prime \prime } +\text{h.c.}\right]\right\},
\end{split}
\end{equation}

\noindent where the $\textbf{M}_i$ with primes are defined similarly to those without primes [c.f. \eqref{Mi}] but with

\begin{subequations}
    \begin{gather}
\Gamma_{l,l^\prime}^{p\mathrm{H}} =-\bra{h_l}z^{p}( \gamma _{1} +\gamma _{2})\ket{h_{l^\prime}}, \\
\begin{split}
    \Gamma_{j,j^\prime}^{p\eta} =&-\frac{1}{3}\bra{+_j}z^{p}( \gamma _{1} +\gamma _{2}) \ket{+_{j^\prime}} \\
    &-\frac{2}{3}\bra{-_j}z^{p}( \gamma _{1} -2\gamma _{2})\ket{-_{j^\prime}},
\end{split} \\
\begin{split}
    T_{j,j^\prime}^{\prime\eta } =&\frac{1}{\sqrt{6} \alpha_{0}}\left(\bra{c_j} z P\ket{+_{j^\prime}} +\bra{+_j} z P \ket{c_{j^\prime}} \right) \\
    &+ i\bra{c_j}\left[ z g/2,k_{z}\right]\ket{c_{j^\prime}} \\
    &-i\left(\bra{+_j} u_{+}^\prime \ket{-_{j^\prime}} -\bra{-_j} u_{-}^\prime\ket{+_{j^\prime}}\right) \\
    &+ \frac{1}{\sqrt{2}}\left[\braket{s_j|-_{j^\prime}} +\braket{-_j|s_{j^\prime}}\right],
\end{split} \\
T_{l,j}^{\prime\text{x}} =\bra{h_l}\left[\frac{z P}{\sqrt{2} \alpha _{0}} \ket{c_j} -\sqrt{3}\left(iu_{+}^\prime\ket{-_j} +\frac{1}{\sqrt{2}} \ket{s_j}\right)\right],\\
\mu_{l,j}^{p} =\frac{\sqrt{3}}{2}\bra{h_l} z^{p}(\gamma_2+\gamma_3) \ket{+_j}, \\
\delta_{l,j}^{p} =\frac{\sqrt{3}}{2}\bra{h_l} z^{p}(\gamma_2-\gamma_3) \ket{+_j}, 
\end{gather}
\end{subequations}

\noindent where $p=1$ corresponds to one prime and $p = 2$ corresponds to two primes. We point out that in basis $\mathcal{B}_0$, the strain components and the SO energy $\Delta$ do not appear explicitly in $\textbf{H}$, since they are already taken in account by the energies $\textbf{E}^{\text{H},\eta}$ and the envelope functions associated with $H_0$. This is because $H$ does not contain any terms such as $\varepsilon_{ii}k_j$ or $\Delta\cdot k_j$.

\color{black}
\section{Perturbative expansion of $\gamma$ and $g_\perp$}\label{sec:pert}

A perturbative expansion of the $k\cdot p$ Hamiltonian for small $\textcolor{\col}{\mathbf{K}}_\parallel$ yields explicit formulas for the effective parameters appearing in \eqref{eff_par} and \eqref{eff_in}. \color{\col}This is obtained by means of a Schrieffer-Wolff transformation (SWT)~\cite{Winkler2003}. To this end, the basis $\mathcal{B}_0$ is convenient since $\textbf{H}$ is exactly diagonal when $\textbf{K}_\parallel$ and $B$ are zero (see Appendix~\ref{sec:model} for the notation). The zero-th order terms in the SWT are directly the energies $\textbf{E}_0$. The first-order terms are given by the diagonal elements in the nonzero blocks of the matrices $\textbf{M}_i$. For instance, the first-order contribution to the effective mass of $\eta$ subbands are the diagonal entries of $\boldsymbol{\Gamma}^\eta$. Similarly, the Rashba parameter $\beta_1 = \alpha_0T_{j,j}^\eta$. The in-plane $g$ factor stems from the $\textbf{M}_1^\prime$ term in \eqref{H_in}, which has a nonvanishing block in the $(\eta +,\eta -)$ subspace, thus yielding $g_\parallel = -2T_{j,j}^{\prime\eta}$ [c.f. \eqref{g_in}]. Besides $\beta_1$ and $g_\parallel$, the effective parameters in \eqref{eff_par} and \eqref{eff_in} require a second- or third-order SWT to be described exactly. \color{black}In particular, $\gamma$ and $g_\perp$ are exactly described by second-order perturbation only. For the $j$-th $\eta$ subband, the terms \textcolor{\col}{$C$ and $D$ in \eqref{pert_eta} correspond to the second-order corrections, and} are given by

\begin{gather}
     \textcolor{\col}{C} = \alpha_0\sum_l{\frac{\left|\textcolor{\col}{T_{l,j}^\text{x}}\right|^2}{E_j^\eta - E_l^\text{H}}}, \\
     \textcolor{\col}{D} = \alpha_0\sum_{j^\prime\neq j}{\frac{|\textcolor{\col}{T_{j,j^\prime}^\eta|^2}}{E_j^\eta - E_{j^\prime}^\eta}}.
\end{gather}

For the $l$-th HH subband, the \textcolor{\col}{$C^\prime$ term in \eqref{pert_HH} is}

\begin{equation}
    \textcolor{\col}{C^\prime} = \alpha_0\sum_j{\frac{\left|\textcolor{\col}{T_{l,j}^\text{x}}\right|^2}{E_l^\text{H} - E_j^\eta}}.
\end{equation}


\begin{thebibliography}{46}%
\makeatletter
\providecommand \@ifxundefined [1]{%
 \@ifx{#1\undefined}
}%
\providecommand \@ifnum [1]{%
 \ifnum #1\expandafter \@firstoftwo
 \else \expandafter \@secondoftwo
 \fi
}%
\providecommand \@ifx [1]{%
 \ifx #1\expandafter \@firstoftwo
 \else \expandafter \@secondoftwo
 \fi
}%
\providecommand \natexlab [1]{#1}%
\providecommand \enquote  [1]{``#1''}%
\providecommand \bibnamefont  [1]{#1}%
\providecommand \bibfnamefont [1]{#1}%
\providecommand \citenamefont [1]{#1}%
\providecommand \href@noop [0]{\@secondoftwo}%
\providecommand \href [0]{\begingroup \@sanitize@url \@href}%
\providecommand \@href[1]{\@@startlink{#1}\@@href}%
\providecommand \@@href[1]{\endgroup#1\@@endlink}%
\providecommand \@sanitize@url [0]{\catcode `\\12\catcode `\$12\catcode
  `\&12\catcode `\#12\catcode `\^12\catcode `\_12\catcode `\%12\relax}%
\providecommand \@@startlink[1]{}%
\providecommand \@@endlink[0]{}%
\providecommand \url  [0]{\begingroup\@sanitize@url \@url }%
\providecommand \@url [1]{\endgroup\@href {#1}{\urlprefix }}%
\providecommand \urlprefix  [0]{URL }%
\providecommand \Eprint [0]{\href }%
\providecommand \doibase [0]{https://doi.org/}%
\providecommand \selectlanguage [0]{\@gobble}%
\providecommand \bibinfo  [0]{\@secondoftwo}%
\providecommand \bibfield  [0]{\@secondoftwo}%
\providecommand \translation [1]{[#1]}%
\providecommand \BibitemOpen [0]{}%
\providecommand \bibitemStop [0]{}%
\providecommand \bibitemNoStop [0]{.\EOS\space}%
\providecommand \EOS [0]{\spacefactor3000\relax}%
\providecommand \BibitemShut  [1]{\csname bibitem#1\endcsname}%
\let\auto@bib@innerbib\@empty
\bibitem [{\citenamefont {Fang}\ \emph {et~al.}(2023)\citenamefont {Fang},
  \citenamefont {Philippopoulos}, \citenamefont {Culcer}, \citenamefont
  {Coish},\ and\ \citenamefont {Chesi}}]{Fang_2023}%
  \BibitemOpen
  \bibfield  {author} {\bibinfo {author} {\bibfnamefont {Y.}~\bibnamefont
  {Fang}}, \bibinfo {author} {\bibfnamefont {P.}~\bibnamefont
  {Philippopoulos}}, \bibinfo {author} {\bibfnamefont {D.}~\bibnamefont
  {Culcer}}, \bibinfo {author} {\bibfnamefont {W.~A.}\ \bibnamefont {Coish}},\
  and\ \bibinfo {author} {\bibfnamefont {S.}~\bibnamefont {Chesi}},\ }\href
  {https://doi.org/10.1088/2633-4356/acb87e} {\bibfield  {journal} {\bibinfo
  {journal} {Materials for Quantum Technology}\ }\textbf {\bibinfo {volume}
  {3}},\ \bibinfo {pages} {012003} (\bibinfo {year} {2023})}\BibitemShut
  {NoStop}%
\bibitem [{\citenamefont {Lawrie}\ \emph {et~al.}(2023)\citenamefont {Lawrie},
  \citenamefont {Rimbach-Russ}, \citenamefont {Riggelen}, \citenamefont
  {Hendrickx}, \citenamefont {Snoo}, \citenamefont {Sammak}, \citenamefont
  {Scappucci}, \citenamefont {Helsen},\ and\ \citenamefont
  {Veldhorst}}]{Lawrie2023}%
  \BibitemOpen
  \bibfield  {author} {\bibinfo {author} {\bibfnamefont {W.~I.~L.}\
  \bibnamefont {Lawrie}}, \bibinfo {author} {\bibfnamefont {M.}~\bibnamefont
  {Rimbach-Russ}}, \bibinfo {author} {\bibfnamefont {F.~v.}\ \bibnamefont
  {Riggelen}}, \bibinfo {author} {\bibfnamefont {N.~W.}\ \bibnamefont
  {Hendrickx}}, \bibinfo {author} {\bibfnamefont {S.~L.~d.}\ \bibnamefont
  {Snoo}}, \bibinfo {author} {\bibfnamefont {A.}~\bibnamefont {Sammak}},
  \bibinfo {author} {\bibfnamefont {G.}~\bibnamefont {Scappucci}}, \bibinfo
  {author} {\bibfnamefont {J.}~\bibnamefont {Helsen}},\ and\ \bibinfo {author}
  {\bibfnamefont {M.}~\bibnamefont {Veldhorst}},\ }\href
  {https://doi.org/10.1038/s41467-023-39334-3} {\bibfield  {journal} {\bibinfo
  {journal} {Nature Communications}\ }\textbf {\bibinfo {volume} {14}},\
  \bibinfo {pages} {3617} (\bibinfo {year} {2023})}\BibitemShut {NoStop}%
\bibitem [{\citenamefont {Jirovec}\ \emph {et~al.}(2021)\citenamefont
  {Jirovec}, \citenamefont {Hofmann}, \citenamefont {Ballabio}, \citenamefont
  {Mutter}, \citenamefont {Tavani}, \citenamefont {Botifoll}, \citenamefont
  {Crippa}, \citenamefont {Kukucka}, \citenamefont {Sagi}, \citenamefont
  {Martins}, \citenamefont {Saez-Mollejo}, \citenamefont {Prieto},
  \citenamefont {Borovkov}, \citenamefont {Arbiol}, \citenamefont {Chrastina},
  \citenamefont {Isella},\ and\ \citenamefont {Katsaros}}]{Jirovec2021}%
  \BibitemOpen
  \bibfield  {author} {\bibinfo {author} {\bibfnamefont {D.}~\bibnamefont
  {Jirovec}}, \bibinfo {author} {\bibfnamefont {A.}~\bibnamefont {Hofmann}},
  \bibinfo {author} {\bibfnamefont {A.}~\bibnamefont {Ballabio}}, \bibinfo
  {author} {\bibfnamefont {P.~M.}\ \bibnamefont {Mutter}}, \bibinfo {author}
  {\bibfnamefont {G.}~\bibnamefont {Tavani}}, \bibinfo {author} {\bibfnamefont
  {M.}~\bibnamefont {Botifoll}}, \bibinfo {author} {\bibfnamefont
  {A.}~\bibnamefont {Crippa}}, \bibinfo {author} {\bibfnamefont
  {J.}~\bibnamefont {Kukucka}}, \bibinfo {author} {\bibfnamefont
  {O.}~\bibnamefont {Sagi}}, \bibinfo {author} {\bibfnamefont {F.}~\bibnamefont
  {Martins}}, \bibinfo {author} {\bibfnamefont {J.}~\bibnamefont
  {Saez-Mollejo}}, \bibinfo {author} {\bibfnamefont {I.}~\bibnamefont
  {Prieto}}, \bibinfo {author} {\bibfnamefont {M.}~\bibnamefont {Borovkov}},
  \bibinfo {author} {\bibfnamefont {J.}~\bibnamefont {Arbiol}}, \bibinfo
  {author} {\bibfnamefont {D.}~\bibnamefont {Chrastina}}, \bibinfo {author}
  {\bibfnamefont {G.}~\bibnamefont {Isella}},\ and\ \bibinfo {author}
  {\bibfnamefont {G.}~\bibnamefont {Katsaros}},\ }\href
  {https://doi.org/10.1038/s41563-021-01022-2} {\bibfield  {journal} {\bibinfo
  {journal} {Nature Materials}\ }\textbf {\bibinfo {volume} {20}},\ \bibinfo
  {pages} {1106} (\bibinfo {year} {2021})}\BibitemShut {NoStop}%
\bibitem [{\citenamefont {Hendrickx}\ \emph {et~al.}(2021)\citenamefont
  {Hendrickx}, \citenamefont {Lawrie}, \citenamefont {Russ}, \citenamefont {van
  Riggelen}, \citenamefont {de~Snoo}, \citenamefont {Schouten}, \citenamefont
  {Sammak}, \citenamefont {Scappucci},\ and\ \citenamefont
  {Veldhorst}}]{Hendrickx2021}%
  \BibitemOpen
  \bibfield  {author} {\bibinfo {author} {\bibfnamefont {N.~W.}\ \bibnamefont
  {Hendrickx}}, \bibinfo {author} {\bibfnamefont {W.~I.~L.}\ \bibnamefont
  {Lawrie}}, \bibinfo {author} {\bibfnamefont {M.}~\bibnamefont {Russ}},
  \bibinfo {author} {\bibfnamefont {F.}~\bibnamefont {van Riggelen}}, \bibinfo
  {author} {\bibfnamefont {S.~L.}\ \bibnamefont {de~Snoo}}, \bibinfo {author}
  {\bibfnamefont {R.~N.}\ \bibnamefont {Schouten}}, \bibinfo {author}
  {\bibfnamefont {A.}~\bibnamefont {Sammak}}, \bibinfo {author} {\bibfnamefont
  {G.}~\bibnamefont {Scappucci}},\ and\ \bibinfo {author} {\bibfnamefont
  {M.}~\bibnamefont {Veldhorst}},\ }\href
  {https://doi.org/10.1038/s41586-021-03332-6} {\bibfield  {journal} {\bibinfo
  {journal} {Nature}\ }\textbf {\bibinfo {volume} {591}},\ \bibinfo {pages}
  {580} (\bibinfo {year} {2021})}\BibitemShut {NoStop}%
\bibitem [{\citenamefont {Hendrickx}\ \emph
  {et~al.}(2020{\natexlab{a}})\citenamefont {Hendrickx}, \citenamefont
  {Lawrie}, \citenamefont {Petit}, \citenamefont {Sammak}, \citenamefont
  {Scappucci},\ and\ \citenamefont {Veldhorst}}]{Hendrickx2020_1}%
  \BibitemOpen
  \bibfield  {author} {\bibinfo {author} {\bibfnamefont {N.~W.}\ \bibnamefont
  {Hendrickx}}, \bibinfo {author} {\bibfnamefont {W.~I.~L.}\ \bibnamefont
  {Lawrie}}, \bibinfo {author} {\bibfnamefont {L.}~\bibnamefont {Petit}},
  \bibinfo {author} {\bibfnamefont {A.}~\bibnamefont {Sammak}}, \bibinfo
  {author} {\bibfnamefont {G.}~\bibnamefont {Scappucci}},\ and\ \bibinfo
  {author} {\bibfnamefont {M.}~\bibnamefont {Veldhorst}},\ }\href
  {https://doi.org/10.1038/s41467-020-17211-7} {\bibfield  {journal} {\bibinfo
  {journal} {Nature Communications}\ }\textbf {\bibinfo {volume} {11}},\
  \bibinfo {pages} {3478} (\bibinfo {year} {2020}{\natexlab{a}})}\BibitemShut
  {NoStop}%
\bibitem [{\citenamefont {Hendrickx}\ \emph
  {et~al.}(2020{\natexlab{b}})\citenamefont {Hendrickx}, \citenamefont
  {Franke}, \citenamefont {Sammak}, \citenamefont {Scappucci},\ and\
  \citenamefont {Veldhorst}}]{Hendrickx2020_2}%
  \BibitemOpen
  \bibfield  {author} {\bibinfo {author} {\bibfnamefont {N.~W.}\ \bibnamefont
  {Hendrickx}}, \bibinfo {author} {\bibfnamefont {D.~P.}\ \bibnamefont
  {Franke}}, \bibinfo {author} {\bibfnamefont {A.}~\bibnamefont {Sammak}},
  \bibinfo {author} {\bibfnamefont {G.}~\bibnamefont {Scappucci}},\ and\
  \bibinfo {author} {\bibfnamefont {M.}~\bibnamefont {Veldhorst}},\ }\href
  {https://doi.org/10.1038/s41586-019-1919-3} {\bibfield  {journal} {\bibinfo
  {journal} {Nature}\ }\textbf {\bibinfo {volume} {577}},\ \bibinfo {pages}
  {487} (\bibinfo {year} {2020}{\natexlab{b}})}\BibitemShut {NoStop}%
\bibitem [{\citenamefont {Lawrie}\ \emph {et~al.}(2020)\citenamefont {Lawrie},
  \citenamefont {Hendrickx}, \citenamefont {van Riggelen}, \citenamefont
  {Russ}, \citenamefont {Petit}, \citenamefont {Sammak}, \citenamefont
  {Scappucci},\ and\ \citenamefont {Veldhorst}}]{Lawrie2020}%
  \BibitemOpen
  \bibfield  {author} {\bibinfo {author} {\bibfnamefont {W.~I.~L.}\
  \bibnamefont {Lawrie}}, \bibinfo {author} {\bibfnamefont {N.~W.}\
  \bibnamefont {Hendrickx}}, \bibinfo {author} {\bibfnamefont {F.}~\bibnamefont
  {van Riggelen}}, \bibinfo {author} {\bibfnamefont {M.}~\bibnamefont {Russ}},
  \bibinfo {author} {\bibfnamefont {L.}~\bibnamefont {Petit}}, \bibinfo
  {author} {\bibfnamefont {A.}~\bibnamefont {Sammak}}, \bibinfo {author}
  {\bibfnamefont {G.}~\bibnamefont {Scappucci}},\ and\ \bibinfo {author}
  {\bibfnamefont {M.}~\bibnamefont {Veldhorst}},\ }\href
  {https://doi.org/10.1021/acs.nanolett.0c02589} {\bibfield  {journal}
  {\bibinfo  {journal} {Nano Letters}\ }\textbf {\bibinfo {volume} {20}},\
  \bibinfo {pages} {7237} (\bibinfo {year} {2020})}\BibitemShut {NoStop}%
\bibitem [{\citenamefont {Assali}\ \emph {et~al.}(2022)\citenamefont {Assali},
  \citenamefont {Attiaoui}, \citenamefont {Del~Vecchio}, \citenamefont
  {Mukherjee}, \citenamefont {Nicolas},\ and\ \citenamefont
  {Moutanabbir}}]{Assali2022}%
  \BibitemOpen
  \bibfield  {author} {\bibinfo {author} {\bibfnamefont {S.}~\bibnamefont
  {Assali}}, \bibinfo {author} {\bibfnamefont {A.}~\bibnamefont {Attiaoui}},
  \bibinfo {author} {\bibfnamefont {P.}~\bibnamefont {Del~Vecchio}}, \bibinfo
  {author} {\bibfnamefont {S.}~\bibnamefont {Mukherjee}}, \bibinfo {author}
  {\bibfnamefont {J.}~\bibnamefont {Nicolas}},\ and\ \bibinfo {author}
  {\bibfnamefont {O.}~\bibnamefont {Moutanabbir}},\ }\href
  {https://doi.org/10.1002/adma.202201192} {\bibfield  {journal} {\bibinfo
  {journal} {Advanced Materials}\ }\textbf {\bibinfo {volume} {34}},\ \bibinfo
  {pages} {2201192} (\bibinfo {year} {2022})}\BibitemShut {NoStop}%
\bibitem [{\citenamefont {Del~Vecchio}\ and\ \citenamefont
  {Moutanabbir}(2023)}]{DelVecchio2023}%
  \BibitemOpen
  \bibfield  {author} {\bibinfo {author} {\bibfnamefont {P.}~\bibnamefont
  {Del~Vecchio}}\ and\ \bibinfo {author} {\bibfnamefont {O.}~\bibnamefont
  {Moutanabbir}},\ }\href {https://doi.org/10.1103/PhysRevB.107.L161406}
  {\bibfield  {journal} {\bibinfo  {journal} {Phys. Rev. B}\ }\textbf {\bibinfo
  {volume} {107}},\ \bibinfo {pages} {L161406} (\bibinfo {year}
  {2023})}\BibitemShut {NoStop}%
\bibitem [{\citenamefont {Moghaddam}\ \emph {et~al.}(2014)\citenamefont
  {Moghaddam}, \citenamefont {Kernreiter}, \citenamefont {Governale},\ and\
  \citenamefont {Z\"ulicke}}]{Moghaddam2014}%
  \BibitemOpen
  \bibfield  {author} {\bibinfo {author} {\bibfnamefont {A.~G.}\ \bibnamefont
  {Moghaddam}}, \bibinfo {author} {\bibfnamefont {T.}~\bibnamefont
  {Kernreiter}}, \bibinfo {author} {\bibfnamefont {M.}~\bibnamefont
  {Governale}},\ and\ \bibinfo {author} {\bibfnamefont {U.}~\bibnamefont
  {Z\"ulicke}},\ }\href {https://doi.org/10.1103/PhysRevB.89.184507} {\bibfield
   {journal} {\bibinfo  {journal} {Phys. Rev. B}\ }\textbf {\bibinfo {volume}
  {89}},\ \bibinfo {pages} {184507} (\bibinfo {year} {2014})}\BibitemShut
  {NoStop}%
\bibitem [{\citenamefont {Moutanabbir}\ \emph {et~al.}(2021)\citenamefont
  {Moutanabbir}, \citenamefont {Assali}, \citenamefont {Gong}, \citenamefont
  {O'Reilly}, \citenamefont {Broderick}, \citenamefont {Marzban}, \citenamefont
  {Witzens}, \citenamefont {Du}, \citenamefont {Yu}, \citenamefont {Chelnokov},
  \citenamefont {Buca},\ and\ \citenamefont {Nam}}]{Moutanabbir2021}%
  \BibitemOpen
  \bibfield  {author} {\bibinfo {author} {\bibfnamefont {O.}~\bibnamefont
  {Moutanabbir}}, \bibinfo {author} {\bibfnamefont {S.}~\bibnamefont {Assali}},
  \bibinfo {author} {\bibfnamefont {X.}~\bibnamefont {Gong}}, \bibinfo {author}
  {\bibfnamefont {E.}~\bibnamefont {O'Reilly}}, \bibinfo {author}
  {\bibfnamefont {C.~A.}\ \bibnamefont {Broderick}}, \bibinfo {author}
  {\bibfnamefont {B.}~\bibnamefont {Marzban}}, \bibinfo {author} {\bibfnamefont
  {J.}~\bibnamefont {Witzens}}, \bibinfo {author} {\bibfnamefont
  {W.}~\bibnamefont {Du}}, \bibinfo {author} {\bibfnamefont {S.-Q.}\
  \bibnamefont {Yu}}, \bibinfo {author} {\bibfnamefont {A.}~\bibnamefont
  {Chelnokov}}, \bibinfo {author} {\bibfnamefont {D.}~\bibnamefont {Buca}},\
  and\ \bibinfo {author} {\bibfnamefont {D.}~\bibnamefont {Nam}},\ }\href
  {https://doi.org/10.1063/5.0043511} {\bibfield  {journal} {\bibinfo
  {journal} {Applied Physics Letters}\ }\textbf {\bibinfo {volume} {118}},\
  \bibinfo {pages} {110502} (\bibinfo {year} {2021})}\BibitemShut {NoStop}%
\bibitem [{\citenamefont {Tai}\ \emph {et~al.}(2021)\citenamefont {Tai},
  \citenamefont {Chiu}, \citenamefont {Liu}, \citenamefont {Kao}, \citenamefont
  {Harris}, \citenamefont {Lu}, \citenamefont {Hsieh}, \citenamefont {Chang},\
  and\ \citenamefont {Li}}]{Tai2021}%
  \BibitemOpen
  \bibfield  {author} {\bibinfo {author} {\bibfnamefont {C.-T.}\ \bibnamefont
  {Tai}}, \bibinfo {author} {\bibfnamefont {P.-Y.}\ \bibnamefont {Chiu}},
  \bibinfo {author} {\bibfnamefont {C.-Y.}\ \bibnamefont {Liu}}, \bibinfo
  {author} {\bibfnamefont {H.-S.}\ \bibnamefont {Kao}}, \bibinfo {author}
  {\bibfnamefont {C.~T.}\ \bibnamefont {Harris}}, \bibinfo {author}
  {\bibfnamefont {T.-M.}\ \bibnamefont {Lu}}, \bibinfo {author} {\bibfnamefont
  {C.-T.}\ \bibnamefont {Hsieh}}, \bibinfo {author} {\bibfnamefont {S.-W.}\
  \bibnamefont {Chang}},\ and\ \bibinfo {author} {\bibfnamefont {J.-Y.}\
  \bibnamefont {Li}},\ }\href
  {https://doi.org/https://doi.org/10.1002/adma.202007862} {\bibfield
  {journal} {\bibinfo  {journal} {Advanced Materials}\ }\textbf {\bibinfo
  {volume} {33}},\ \bibinfo {pages} {2007862} (\bibinfo {year}
  {2021})}\BibitemShut {NoStop}%
\bibitem [{\citenamefont {Ferrari}\ \emph {et~al.}(2023)\citenamefont
  {Ferrari}, \citenamefont {Marcantonio}, \citenamefont {Murphy-Armando},
  \citenamefont {Virgilio},\ and\ \citenamefont {Pezzoli}}]{Ferrari2023}%
  \BibitemOpen
  \bibfield  {author} {\bibinfo {author} {\bibfnamefont {B.~M.}\ \bibnamefont
  {Ferrari}}, \bibinfo {author} {\bibfnamefont {F.}~\bibnamefont
  {Marcantonio}}, \bibinfo {author} {\bibfnamefont {F.}~\bibnamefont
  {Murphy-Armando}}, \bibinfo {author} {\bibfnamefont {M.}~\bibnamefont
  {Virgilio}},\ and\ \bibinfo {author} {\bibfnamefont {F.}~\bibnamefont
  {Pezzoli}},\ }\href {https://doi.org/10.1103/PhysRevResearch.5.L022035}
  {\bibfield  {journal} {\bibinfo  {journal} {Phys. Rev. Res.}\ }\textbf
  {\bibinfo {volume} {5}},\ \bibinfo {pages} {L022035} (\bibinfo {year}
  {2023})}\BibitemShut {NoStop}%
\bibitem [{\citenamefont {Fettu}\ \emph {et~al.}(2023)\citenamefont {Fettu},
  \citenamefont {Sipe},\ and\ \citenamefont {Moutanabbir}}]{Fettu2023}%
  \BibitemOpen
  \bibfield  {author} {\bibinfo {author} {\bibfnamefont {G.}~\bibnamefont
  {Fettu}}, \bibinfo {author} {\bibfnamefont {J.~E.}\ \bibnamefont {Sipe}},\
  and\ \bibinfo {author} {\bibfnamefont {O.}~\bibnamefont {Moutanabbir}},\
  }\href {https://doi.org/10.1103/PhysRevB.107.165202} {\bibfield  {journal}
  {\bibinfo  {journal} {Phys. Rev. B}\ }\textbf {\bibinfo {volume} {107}},\
  \bibinfo {pages} {165202} (\bibinfo {year} {2023})}\BibitemShut {NoStop}%
\bibitem [{\citenamefont {Van~de Walle}(1989)}]{VandeWalle1989}%
  \BibitemOpen
  \bibfield  {author} {\bibinfo {author} {\bibfnamefont {C.~G.}\ \bibnamefont
  {Van~de Walle}},\ }\href {https://doi.org/10.1103/PhysRevB.39.1871}
  {\bibfield  {journal} {\bibinfo  {journal} {Phys. Rev. B}\ }\textbf {\bibinfo
  {volume} {39}},\ \bibinfo {pages} {1871} (\bibinfo {year}
  {1989})}\BibitemShut {NoStop}%
\bibitem [{\citenamefont {People}\ and\ \citenamefont
  {Bean}(1985)}]{People1985}%
  \BibitemOpen
  \bibfield  {author} {\bibinfo {author} {\bibfnamefont {R.}~\bibnamefont
  {People}}\ and\ \bibinfo {author} {\bibfnamefont {J.~C.}\ \bibnamefont
  {Bean}},\ }\href {https://doi.org/10.1063/1.96206} {\bibfield  {journal}
  {\bibinfo  {journal} {Applied Physics Letters}\ }\textbf {\bibinfo {volume}
  {47}},\ \bibinfo {pages} {322} (\bibinfo {year} {1985})}\BibitemShut
  {NoStop}%
\bibitem [{\citenamefont {People}\ and\ \citenamefont
  {Bean}(1986)}]{People1986E}%
  \BibitemOpen
  \bibfield  {author} {\bibinfo {author} {\bibfnamefont {R.}~\bibnamefont
  {People}}\ and\ \bibinfo {author} {\bibfnamefont {J.~C.}\ \bibnamefont
  {Bean}},\ }\href {https://doi.org/10.1063/1.97637} {\bibfield  {journal}
  {\bibinfo  {journal} {Applied Physics Letters}\ }\textbf {\bibinfo {volume}
  {49}},\ \bibinfo {pages} {229} (\bibinfo {year} {1986})}\BibitemShut
  {NoStop}%
\bibitem [{\citenamefont {Wang}\ \emph {et~al.}(2021)\citenamefont {Wang},
  \citenamefont {Marcellina}, \citenamefont {Hamilton}, \citenamefont {Cullen},
  \citenamefont {Rogge}, \citenamefont {Salfi},\ and\ \citenamefont
  {Culcer}}]{Wang2021}%
  \BibitemOpen
  \bibfield  {author} {\bibinfo {author} {\bibfnamefont {Z.}~\bibnamefont
  {Wang}}, \bibinfo {author} {\bibfnamefont {E.}~\bibnamefont {Marcellina}},
  \bibinfo {author} {\bibfnamefont {A.~R.}\ \bibnamefont {Hamilton}}, \bibinfo
  {author} {\bibfnamefont {J.~H.}\ \bibnamefont {Cullen}}, \bibinfo {author}
  {\bibfnamefont {S.}~\bibnamefont {Rogge}}, \bibinfo {author} {\bibfnamefont
  {J.}~\bibnamefont {Salfi}},\ and\ \bibinfo {author} {\bibfnamefont
  {D.}~\bibnamefont {Culcer}},\ }\href
  {https://doi.org/10.1038/s41534-021-00386-2} {\bibfield  {journal} {\bibinfo
  {journal} {npj Quantum Information}\ }\textbf {\bibinfo {volume} {7}},\
  \bibinfo {pages} {54} (\bibinfo {year} {2021})}\BibitemShut {NoStop}%
\bibitem [{\citenamefont {Wang}\ \emph {et~al.}(2022)\citenamefont {Wang},
  \citenamefont {Scappucci}, \citenamefont {Veldhorst},\ and\ \citenamefont
  {Russ}}]{Wang2022arxiv}%
  \BibitemOpen
  \bibfield  {author} {\bibinfo {author} {\bibfnamefont {C.-A.}\ \bibnamefont
  {Wang}}, \bibinfo {author} {\bibfnamefont {G.}~\bibnamefont {Scappucci}},
  \bibinfo {author} {\bibfnamefont {M.}~\bibnamefont {Veldhorst}},\ and\
  \bibinfo {author} {\bibfnamefont {M.}~\bibnamefont {Russ}},\ }\href@noop {}
  {\bibinfo {title} {{Modelling of planar germanium hole qubits in electric and
  magnetic fields}}} (\bibinfo {year} {2022}),\ \Eprint
  {https://arxiv.org/abs/2208.04795} {arXiv:2208.04795 [cond-mat.mes-hall]}
  \BibitemShut {NoStop}%
\bibitem [{\citenamefont {Hayden}\ \emph {et~al.}(1991)\citenamefont {Hayden},
  \citenamefont {Maude}, \citenamefont {Eaves}, \citenamefont {Valadares},
  \citenamefont {Henini}, \citenamefont {Sheard}, \citenamefont {Hughes},
  \citenamefont {Portal},\ and\ \citenamefont {Cury}}]{Hayden1991}%
  \BibitemOpen
  \bibfield  {author} {\bibinfo {author} {\bibfnamefont {R.~K.}\ \bibnamefont
  {Hayden}}, \bibinfo {author} {\bibfnamefont {D.~K.}\ \bibnamefont {Maude}},
  \bibinfo {author} {\bibfnamefont {L.}~\bibnamefont {Eaves}}, \bibinfo
  {author} {\bibfnamefont {E.~C.}\ \bibnamefont {Valadares}}, \bibinfo {author}
  {\bibfnamefont {M.}~\bibnamefont {Henini}}, \bibinfo {author} {\bibfnamefont
  {F.~W.}\ \bibnamefont {Sheard}}, \bibinfo {author} {\bibfnamefont {O.~H.}\
  \bibnamefont {Hughes}}, \bibinfo {author} {\bibfnamefont {J.~C.}\
  \bibnamefont {Portal}},\ and\ \bibinfo {author} {\bibfnamefont
  {L.}~\bibnamefont {Cury}},\ }\href
  {https://doi.org/10.1103/PhysRevLett.66.1749} {\bibfield  {journal} {\bibinfo
   {journal} {Phys. Rev. Lett.}\ }\textbf {\bibinfo {volume} {66}},\ \bibinfo
  {pages} {1749} (\bibinfo {year} {1991})}\BibitemShut {NoStop}%
\bibitem [{\citenamefont {Jirovec}\ \emph {et~al.}(2022)\citenamefont
  {Jirovec}, \citenamefont {Mutter}, \citenamefont {Hofmann}, \citenamefont
  {Crippa}, \citenamefont {Rychetsky}, \citenamefont {Craig}, \citenamefont
  {Kukucka}, \citenamefont {Martins}, \citenamefont {Ballabio}, \citenamefont
  {Ares}, \citenamefont {Chrastina}, \citenamefont {Isella}, \citenamefont
  {Burkard},\ and\ \citenamefont {Katsaros}}]{Jirovec2022}%
  \BibitemOpen
  \bibfield  {author} {\bibinfo {author} {\bibfnamefont {D.}~\bibnamefont
  {Jirovec}}, \bibinfo {author} {\bibfnamefont {P.~M.}\ \bibnamefont {Mutter}},
  \bibinfo {author} {\bibfnamefont {A.}~\bibnamefont {Hofmann}}, \bibinfo
  {author} {\bibfnamefont {A.}~\bibnamefont {Crippa}}, \bibinfo {author}
  {\bibfnamefont {M.}~\bibnamefont {Rychetsky}}, \bibinfo {author}
  {\bibfnamefont {D.~L.}\ \bibnamefont {Craig}}, \bibinfo {author}
  {\bibfnamefont {J.}~\bibnamefont {Kukucka}}, \bibinfo {author} {\bibfnamefont
  {F.}~\bibnamefont {Martins}}, \bibinfo {author} {\bibfnamefont
  {A.}~\bibnamefont {Ballabio}}, \bibinfo {author} {\bibfnamefont
  {N.}~\bibnamefont {Ares}}, \bibinfo {author} {\bibfnamefont {D.}~\bibnamefont
  {Chrastina}}, \bibinfo {author} {\bibfnamefont {G.}~\bibnamefont {Isella}},
  \bibinfo {author} {\bibfnamefont {G.}~\bibnamefont {Burkard}},\ and\ \bibinfo
  {author} {\bibfnamefont {G.}~\bibnamefont {Katsaros}},\ }\href
  {https://doi.org/10.1103/PhysRevLett.128.126803} {\bibfield  {journal}
  {\bibinfo  {journal} {Phys. Rev. Lett.}\ }\textbf {\bibinfo {volume} {128}},\
  \bibinfo {pages} {126803} (\bibinfo {year} {2022})}\BibitemShut {NoStop}%
\bibitem [{\citenamefont {Watzinger}\ \emph {et~al.}(2018)\citenamefont
  {Watzinger}, \citenamefont {Kuku{\v{c}}ka}, \citenamefont
  {Vuku{\v{s}}i{\'{c}}}, \citenamefont {Gao}, \citenamefont {Wang},
  \citenamefont {Sch{\"a}ffler}, \citenamefont {Zhang},\ and\ \citenamefont
  {Katsaros}}]{Watzinger2018}%
  \BibitemOpen
  \bibfield  {author} {\bibinfo {author} {\bibfnamefont {H.}~\bibnamefont
  {Watzinger}}, \bibinfo {author} {\bibfnamefont {J.}~\bibnamefont
  {Kuku{\v{c}}ka}}, \bibinfo {author} {\bibfnamefont {L.}~\bibnamefont
  {Vuku{\v{s}}i{\'{c}}}}, \bibinfo {author} {\bibfnamefont {F.}~\bibnamefont
  {Gao}}, \bibinfo {author} {\bibfnamefont {T.}~\bibnamefont {Wang}}, \bibinfo
  {author} {\bibfnamefont {F.}~\bibnamefont {Sch{\"a}ffler}}, \bibinfo {author}
  {\bibfnamefont {J.-J.}\ \bibnamefont {Zhang}},\ and\ \bibinfo {author}
  {\bibfnamefont {G.}~\bibnamefont {Katsaros}},\ }\href
  {https://doi.org/10.1038/s41467-018-06418-4} {\bibfield  {journal} {\bibinfo
  {journal} {Nature Communications}\ }\textbf {\bibinfo {volume} {9}},\
  \bibinfo {pages} {3902} (\bibinfo {year} {2018})}\BibitemShut {NoStop}%
\bibitem [{\citenamefont {Hendrickx}\ \emph {et~al.}(2018)\citenamefont
  {Hendrickx}, \citenamefont {Franke}, \citenamefont {Sammak}, \citenamefont
  {Kouwenhoven}, \citenamefont {Sabbagh}, \citenamefont {Yeoh}, \citenamefont
  {Li}, \citenamefont {Tagliaferri}, \citenamefont {Virgilio}, \citenamefont
  {Capellini}, \citenamefont {Scappucci},\ and\ \citenamefont
  {Veldhorst}}]{Hendrickx2018}%
  \BibitemOpen
  \bibfield  {author} {\bibinfo {author} {\bibfnamefont {N.~W.}\ \bibnamefont
  {Hendrickx}}, \bibinfo {author} {\bibfnamefont {D.~P.}\ \bibnamefont
  {Franke}}, \bibinfo {author} {\bibfnamefont {A.}~\bibnamefont {Sammak}},
  \bibinfo {author} {\bibfnamefont {M.}~\bibnamefont {Kouwenhoven}}, \bibinfo
  {author} {\bibfnamefont {D.}~\bibnamefont {Sabbagh}}, \bibinfo {author}
  {\bibfnamefont {L.}~\bibnamefont {Yeoh}}, \bibinfo {author} {\bibfnamefont
  {R.}~\bibnamefont {Li}}, \bibinfo {author} {\bibfnamefont {M.~L.~V.}\
  \bibnamefont {Tagliaferri}}, \bibinfo {author} {\bibfnamefont
  {M.}~\bibnamefont {Virgilio}}, \bibinfo {author} {\bibfnamefont
  {G.}~\bibnamefont {Capellini}}, \bibinfo {author} {\bibfnamefont
  {G.}~\bibnamefont {Scappucci}},\ and\ \bibinfo {author} {\bibfnamefont
  {M.}~\bibnamefont {Veldhorst}},\ }\href
  {https://doi.org/10.1038/s41467-018-05299-x} {\bibfield  {journal} {\bibinfo
  {journal} {Nature Communications}\ }\textbf {\bibinfo {volume} {9}},\
  \bibinfo {pages} {2835} (\bibinfo {year} {2018})}\BibitemShut {NoStop}%
\bibitem [{\citenamefont {Drichko}\ \emph {et~al.}(2018)\citenamefont
  {Drichko}, \citenamefont {Dmitriev}, \citenamefont {Malysh}, \citenamefont
  {Smirnov}, \citenamefont {von Känel}, \citenamefont {Kummer}, \citenamefont
  {Chrastina},\ and\ \citenamefont {Isella}}]{Drichko2018}%
  \BibitemOpen
  \bibfield  {author} {\bibinfo {author} {\bibfnamefont {I.~L.}\ \bibnamefont
  {Drichko}}, \bibinfo {author} {\bibfnamefont {A.~A.}\ \bibnamefont
  {Dmitriev}}, \bibinfo {author} {\bibfnamefont {V.~A.}\ \bibnamefont
  {Malysh}}, \bibinfo {author} {\bibfnamefont {I.~Y.}\ \bibnamefont {Smirnov}},
  \bibinfo {author} {\bibfnamefont {H.}~\bibnamefont {von Känel}}, \bibinfo
  {author} {\bibfnamefont {M.}~\bibnamefont {Kummer}}, \bibinfo {author}
  {\bibfnamefont {D.}~\bibnamefont {Chrastina}},\ and\ \bibinfo {author}
  {\bibfnamefont {G.}~\bibnamefont {Isella}},\ }\href
  {https://doi.org/10.1063/1.5025413} {\bibfield  {journal} {\bibinfo
  {journal} {Journal of Applied Physics}\ }\textbf {\bibinfo {volume} {123}},\
  \bibinfo {pages} {165703} (\bibinfo {year} {2018})}\BibitemShut {NoStop}%
\bibitem [{\citenamefont {Wimbauer}\ \emph {et~al.}(1994)\citenamefont
  {Wimbauer}, \citenamefont {Oettinger}, \citenamefont {Efros}, \citenamefont
  {Meyer},\ and\ \citenamefont {Brugger}}]{Wimbauer1994}%
  \BibitemOpen
  \bibfield  {author} {\bibinfo {author} {\bibfnamefont {T.}~\bibnamefont
  {Wimbauer}}, \bibinfo {author} {\bibfnamefont {K.}~\bibnamefont {Oettinger}},
  \bibinfo {author} {\bibfnamefont {A.~L.}\ \bibnamefont {Efros}}, \bibinfo
  {author} {\bibfnamefont {B.~K.}\ \bibnamefont {Meyer}},\ and\ \bibinfo
  {author} {\bibfnamefont {H.}~\bibnamefont {Brugger}},\ }\href
  {https://doi.org/10.1103/PhysRevB.50.8889} {\bibfield  {journal} {\bibinfo
  {journal} {Phys. Rev. B}\ }\textbf {\bibinfo {volume} {50}},\ \bibinfo
  {pages} {8889} (\bibinfo {year} {1994})}\BibitemShut {NoStop}%
\bibitem [{\citenamefont {Reeber}\ and\ \citenamefont
  {Wang}(1996)}]{Reeber1996}%
  \BibitemOpen
  \bibfield  {author} {\bibinfo {author} {\bibfnamefont {R.~R.}\ \bibnamefont
  {Reeber}}\ and\ \bibinfo {author} {\bibfnamefont {K.}~\bibnamefont {Wang}},\
  }\href {https://doi.org/https://doi.org/10.1016/S0254-0584(96)01808-1}
  {\bibfield  {journal} {\bibinfo  {journal} {Materials Chemistry and Physics}\
  }\textbf {\bibinfo {volume} {46}},\ \bibinfo {pages} {259 } (\bibinfo {year}
  {1996})}\BibitemShut {NoStop}%
\bibitem [{\citenamefont {Madelung}(1991)}]{Madelung1991}%
  \BibitemOpen
  \bibinfo {editor} {\bibfnamefont {O.}~\bibnamefont {Madelung}},\ ed.,\ \href
  {https://doi.org/10.1007/978-3-642-45681-7} {\emph {\bibinfo {title}
  {{Semiconductors, Group IV Elements and III-V Compounds}}}}\ (\bibinfo
  {publisher} {Springer-Verlag Berlin Heidelberg},\ \bibinfo {year}
  {1991})\BibitemShut {NoStop}%
\bibitem [{\citenamefont {Polak}\ \emph {et~al.}(2017)\citenamefont {Polak},
  \citenamefont {Scharoch},\ and\ \citenamefont {Kudrawiec}}]{Polak2017}%
  \BibitemOpen
  \bibfield  {author} {\bibinfo {author} {\bibfnamefont {M.~P.}\ \bibnamefont
  {Polak}}, \bibinfo {author} {\bibfnamefont {P.}~\bibnamefont {Scharoch}},\
  and\ \bibinfo {author} {\bibfnamefont {R.}~\bibnamefont {Kudrawiec}},\ }\href
  {https://doi.org/10.1088/1361-6463/aa67bf} {\bibfield  {journal} {\bibinfo
  {journal} {Journal of Physics D: Applied Physics}\ }\textbf {\bibinfo
  {volume} {50}},\ \bibinfo {pages} {195103} (\bibinfo {year}
  {2017})}\BibitemShut {NoStop}%
\bibitem [{\citenamefont {Bertrand}\ \emph {et~al.}(2019)\citenamefont
  {Bertrand}, \citenamefont {Thai}, \citenamefont {Chrétien}, \citenamefont
  {Pauc}, \citenamefont {Aubin}, \citenamefont {Milord}, \citenamefont
  {Gassenq}, \citenamefont {Hartmann}, \citenamefont {Chelnokov}, \citenamefont
  {Calvo},\ and\ \citenamefont {Reboud}}]{Bertrand2019}%
  \BibitemOpen
  \bibfield  {author} {\bibinfo {author} {\bibfnamefont {M.}~\bibnamefont
  {Bertrand}}, \bibinfo {author} {\bibfnamefont {Q.-M.}\ \bibnamefont {Thai}},
  \bibinfo {author} {\bibfnamefont {J.}~\bibnamefont {Chrétien}}, \bibinfo
  {author} {\bibfnamefont {N.}~\bibnamefont {Pauc}}, \bibinfo {author}
  {\bibfnamefont {J.}~\bibnamefont {Aubin}}, \bibinfo {author} {\bibfnamefont
  {L.}~\bibnamefont {Milord}}, \bibinfo {author} {\bibfnamefont
  {A.}~\bibnamefont {Gassenq}}, \bibinfo {author} {\bibfnamefont {J.-M.}\
  \bibnamefont {Hartmann}}, \bibinfo {author} {\bibfnamefont {A.}~\bibnamefont
  {Chelnokov}}, \bibinfo {author} {\bibfnamefont {V.}~\bibnamefont {Calvo}},\
  and\ \bibinfo {author} {\bibfnamefont {V.}~\bibnamefont {Reboud}},\ }\href
  {https://doi.org/10.1002/andp.201800396} {\bibfield  {journal} {\bibinfo
  {journal} {Annalen der Physik}\ }\textbf {\bibinfo {volume} {531}},\ \bibinfo
  {pages} {1800396} (\bibinfo {year} {2019})}\BibitemShut {NoStop}%
\bibitem [{\citenamefont {Varshni}(1967)}]{Varshni1967}%
  \BibitemOpen
  \bibfield  {author} {\bibinfo {author} {\bibfnamefont {Y.}~\bibnamefont
  {Varshni}},\ }\href
  {https://doi.org/https://doi.org/10.1016/0031-8914(67)90062-6} {\bibfield
  {journal} {\bibinfo  {journal} {Physica}\ }\textbf {\bibinfo {volume} {34}},\
  \bibinfo {pages} {149} (\bibinfo {year} {1967})}\BibitemShut {NoStop}%
\bibitem [{\citenamefont {Lawaetz}(1971)}]{Lawaetz1971}%
  \BibitemOpen
  \bibfield  {author} {\bibinfo {author} {\bibfnamefont {P.}~\bibnamefont
  {Lawaetz}},\ }\href {https://doi.org/10.1103/PhysRevB.4.3460} {\bibfield
  {journal} {\bibinfo  {journal} {Phys. Rev. B}\ }\textbf {\bibinfo {volume}
  {4}},\ \bibinfo {pages} {3460} (\bibinfo {year} {1971})}\BibitemShut
  {NoStop}%
\bibitem [{\citenamefont {Men\'endez}\ and\ \citenamefont
  {Kouvetakis}(2004)}]{Menendez2004}%
  \BibitemOpen
  \bibfield  {author} {\bibinfo {author} {\bibfnamefont {J.}~\bibnamefont
  {Men\'endez}}\ and\ \bibinfo {author} {\bibfnamefont {J.}~\bibnamefont
  {Kouvetakis}},\ }\href {https://doi.org/10.1063/1.1784032} {\bibfield
  {journal} {\bibinfo  {journal} {Applied Physics Letters}\ }\textbf {\bibinfo
  {volume} {85}},\ \bibinfo {pages} {1175} (\bibinfo {year}
  {2004})}\BibitemShut {NoStop}%
\bibitem [{\citenamefont {Van~de Walle}\ and\ \citenamefont
  {Martin}(1986)}]{VandeWalle1986}%
  \BibitemOpen
  \bibfield  {author} {\bibinfo {author} {\bibfnamefont {C.~G.}\ \bibnamefont
  {Van~de Walle}}\ and\ \bibinfo {author} {\bibfnamefont {R.~M.}\ \bibnamefont
  {Martin}},\ }\href {https://doi.org/10.1103/PhysRevB.34.5621} {\bibfield
  {journal} {\bibinfo  {journal} {Phys. Rev. B}\ }\textbf {\bibinfo {volume}
  {34}},\ \bibinfo {pages} {5621} (\bibinfo {year} {1986})}\BibitemShut
  {NoStop}%
\bibitem [{\citenamefont {Winkler}(2003)}]{Winkler2003}%
  \BibitemOpen
  \bibfield  {author} {\bibinfo {author} {\bibfnamefont {R.}~\bibnamefont
  {Winkler}},\ }\href {https://doi.org/10.1007/b13586} {\emph {\bibinfo {title}
  {{Spin-orbit coupling effects in two-dimensional electron and hole
  systems}}}},\ Springer tracts in modern physics\ (\bibinfo  {publisher}
  {Springer},\ \bibinfo {address} {Berlin},\ \bibinfo {year}
  {2003})\BibitemShut {NoStop}%
\bibitem [{\citenamefont {D'Costa}\ \emph {et~al.}(2006)\citenamefont
  {D'Costa}, \citenamefont {Cook}, \citenamefont {Birdwell}, \citenamefont
  {Littler}, \citenamefont {Canonico}, \citenamefont {Zollner}, \citenamefont
  {Kouvetakis},\ and\ \citenamefont {Men\'endez}}]{DCosta2006}%
  \BibitemOpen
  \bibfield  {author} {\bibinfo {author} {\bibfnamefont {V.~R.}\ \bibnamefont
  {D'Costa}}, \bibinfo {author} {\bibfnamefont {C.~S.}\ \bibnamefont {Cook}},
  \bibinfo {author} {\bibfnamefont {A.~G.}\ \bibnamefont {Birdwell}}, \bibinfo
  {author} {\bibfnamefont {C.~L.}\ \bibnamefont {Littler}}, \bibinfo {author}
  {\bibfnamefont {M.}~\bibnamefont {Canonico}}, \bibinfo {author}
  {\bibfnamefont {S.}~\bibnamefont {Zollner}}, \bibinfo {author} {\bibfnamefont
  {J.}~\bibnamefont {Kouvetakis}},\ and\ \bibinfo {author} {\bibfnamefont
  {J.}~\bibnamefont {Men\'endez}},\ }\href
  {https://doi.org/10.1103/PhysRevB.73.125207} {\bibfield  {journal} {\bibinfo
  {journal} {Phys. Rev. B}\ }\textbf {\bibinfo {volume} {73}},\ \bibinfo
  {pages} {125207} (\bibinfo {year} {2006})}\BibitemShut {NoStop}%
\bibitem [{\citenamefont {Li}\ \emph {et~al.}(2006)\citenamefont {Li},
  \citenamefont {Gong},\ and\ \citenamefont {Wei}}]{Li2006}%
  \BibitemOpen
  \bibfield  {author} {\bibinfo {author} {\bibfnamefont {Y.-H.}\ \bibnamefont
  {Li}}, \bibinfo {author} {\bibfnamefont {X.~G.}\ \bibnamefont {Gong}},\ and\
  \bibinfo {author} {\bibfnamefont {S.-H.}\ \bibnamefont {Wei}},\ }\href
  {https://doi.org/10.1103/PhysRevB.73.245206} {\bibfield  {journal} {\bibinfo
  {journal} {Phys. Rev. B}\ }\textbf {\bibinfo {volume} {73}},\ \bibinfo
  {pages} {245206} (\bibinfo {year} {2006})}\BibitemShut {NoStop}%
\bibitem [{\citenamefont {Brudevoll}\ \emph {et~al.}(1993)\citenamefont
  {Brudevoll}, \citenamefont {Citrin}, \citenamefont {Cardona},\ and\
  \citenamefont {Christensen}}]{Brudevoll1993}%
  \BibitemOpen
  \bibfield  {author} {\bibinfo {author} {\bibfnamefont {T.}~\bibnamefont
  {Brudevoll}}, \bibinfo {author} {\bibfnamefont {D.~S.}\ \bibnamefont
  {Citrin}}, \bibinfo {author} {\bibfnamefont {M.}~\bibnamefont {Cardona}},\
  and\ \bibinfo {author} {\bibfnamefont {N.~E.}\ \bibnamefont {Christensen}},\
  }\href {https://doi.org/10.1103/PhysRevB.48.8629} {\bibfield  {journal}
  {\bibinfo  {journal} {Phys. Rev. B}\ }\textbf {\bibinfo {volume} {48}},\
  \bibinfo {pages} {8629} (\bibinfo {year} {1993})}\BibitemShut {NoStop}%
\bibitem [{\citenamefont {Willatzen}\ \emph {et~al.}(1995)\citenamefont
  {Willatzen}, \citenamefont {Lew Yan~Voon}, \citenamefont {Santos},
  \citenamefont {Cardona}, \citenamefont {Munzar},\ and\ \citenamefont
  {Christensen}}]{Willatzen1995}%
  \BibitemOpen
  \bibfield  {author} {\bibinfo {author} {\bibfnamefont {M.}~\bibnamefont
  {Willatzen}}, \bibinfo {author} {\bibfnamefont {L.~C.}\ \bibnamefont {Lew
  Yan~Voon}}, \bibinfo {author} {\bibfnamefont {P.~V.}\ \bibnamefont {Santos}},
  \bibinfo {author} {\bibfnamefont {M.}~\bibnamefont {Cardona}}, \bibinfo
  {author} {\bibfnamefont {D.}~\bibnamefont {Munzar}},\ and\ \bibinfo {author}
  {\bibfnamefont {N.~E.}\ \bibnamefont {Christensen}},\ }\href
  {https://doi.org/10.1103/PhysRevB.52.5070} {\bibfield  {journal} {\bibinfo
  {journal} {Phys. Rev. B}\ }\textbf {\bibinfo {volume} {52}},\ \bibinfo
  {pages} {5070} (\bibinfo {year} {1995})}\BibitemShut {NoStop}%
\bibitem [{\citenamefont {Weber}\ and\ \citenamefont
  {Alonso}(1989)}]{Weber1989}%
  \BibitemOpen
  \bibfield  {author} {\bibinfo {author} {\bibfnamefont {J.}~\bibnamefont
  {Weber}}\ and\ \bibinfo {author} {\bibfnamefont {M.~I.}\ \bibnamefont
  {Alonso}},\ }\href {https://doi.org/10.1103/PhysRevB.40.5683} {\bibfield
  {journal} {\bibinfo  {journal} {Phys. Rev. B}\ }\textbf {\bibinfo {volume}
  {40}},\ \bibinfo {pages} {5683} (\bibinfo {year} {1989})}\BibitemShut
  {NoStop}%
\bibitem [{\citenamefont {Lu~Low}\ \emph {et~al.}(2012)\citenamefont {Lu~Low},
  \citenamefont {Yang}, \citenamefont {Han}, \citenamefont {Fan},\ and\
  \citenamefont {Yeo}}]{LuLow2012}%
  \BibitemOpen
  \bibfield  {author} {\bibinfo {author} {\bibfnamefont {K.}~\bibnamefont
  {Lu~Low}}, \bibinfo {author} {\bibfnamefont {Y.}~\bibnamefont {Yang}},
  \bibinfo {author} {\bibfnamefont {G.}~\bibnamefont {Han}}, \bibinfo {author}
  {\bibfnamefont {W.}~\bibnamefont {Fan}},\ and\ \bibinfo {author}
  {\bibfnamefont {Y.-C.}\ \bibnamefont {Yeo}},\ }\href
  {https://doi.org/10.1063/1.4767381} {\bibfield  {journal} {\bibinfo
  {journal} {Journal of Applied Physics}\ }\textbf {\bibinfo {volume} {112}},\
  \bibinfo {pages} {103715} (\bibinfo {year} {2012})}\BibitemShut {NoStop}%
\bibitem [{\citenamefont {Foreman}(1997)}]{Foreman1997}%
  \BibitemOpen
  \bibfield  {author} {\bibinfo {author} {\bibfnamefont {B.~A.}\ \bibnamefont
  {Foreman}},\ }\href {https://doi.org/10.1103/PhysRevB.56.R12748} {\bibfield
  {journal} {\bibinfo  {journal} {Phys. Rev. B}\ }\textbf {\bibinfo {volume}
  {56}},\ \bibinfo {pages} {R12748} (\bibinfo {year} {1997})}\BibitemShut
  {NoStop}%
\bibitem [{\citenamefont {Veprek}\ \emph {et~al.}(2007)\citenamefont {Veprek},
  \citenamefont {Steiger},\ and\ \citenamefont {Witzigmann}}]{Veprek2007}%
  \BibitemOpen
  \bibfield  {author} {\bibinfo {author} {\bibfnamefont {R.~G.}\ \bibnamefont
  {Veprek}}, \bibinfo {author} {\bibfnamefont {S.}~\bibnamefont {Steiger}},\
  and\ \bibinfo {author} {\bibfnamefont {B.}~\bibnamefont {Witzigmann}},\
  }\href {https://doi.org/10.1103/PhysRevB.76.165320} {\bibfield  {journal}
  {\bibinfo  {journal} {Phys. Rev. B}\ }\textbf {\bibinfo {volume} {76}},\
  \bibinfo {pages} {165320} (\bibinfo {year} {2007})}\BibitemShut {NoStop}%
\bibitem [{\citenamefont {Ei{\ss}feller}\ and\ \citenamefont
  {Vogl}(2011)}]{Eissfeller2011}%
  \BibitemOpen
  \bibfield  {author} {\bibinfo {author} {\bibfnamefont {T.}~\bibnamefont
  {Ei{\ss}feller}}\ and\ \bibinfo {author} {\bibfnamefont {P.}~\bibnamefont
  {Vogl}},\ }\href {https://doi.org/10.1103/PhysRevB.84.195122} {\bibfield
  {journal} {\bibinfo  {journal} {Phys. Rev. B}\ }\textbf {\bibinfo {volume}
  {84}},\ \bibinfo {pages} {195122} (\bibinfo {year} {2011})}\BibitemShut
  {NoStop}%
\bibitem [{\citenamefont {Ei{\ss}feller}(2012)}]{Eissfeller2012}%
  \BibitemOpen
  \bibfield  {author} {\bibinfo {author} {\bibfnamefont {T.}~\bibnamefont
  {Ei{\ss}feller}},\ }\emph {\bibinfo {title} {{Theory of the electronic
  structure of quantum dots in external fields}}},\ \href
  {https://www.proquest.com/dissertations-theses/theory-electronic-structure-quantum-dots-external/docview/2193178110/se-2?accountid=40695}
  {Ph.D. thesis},\ \bibinfo  {school} {Technische Universitaet Muenchen
  (Germany)} (\bibinfo {year} {2012})\BibitemShut {NoStop}%
\bibitem [{\citenamefont {Voon}\ and\ \citenamefont
  {Willatzen}(2009)}]{Voon2009}%
  \BibitemOpen
  \bibfield  {author} {\bibinfo {author} {\bibfnamefont {L.}~\bibnamefont
  {Voon}}\ and\ \bibinfo {author} {\bibfnamefont {M.}~\bibnamefont
  {Willatzen}},\ }\href {https://books.google.co.cr/books?id=t46OZQrEd8QC}
  {\emph {\bibinfo {title} {{The k p Method: Electronic Properties of
  Semiconductors}}}}\ (\bibinfo  {publisher} {Springer Berlin Heidelberg},\
  \bibinfo {year} {2009})\BibitemShut {NoStop}%
\bibitem [{\citenamefont {Bir}\ and\ \citenamefont {Pikus}(1974)}]{Bir1974}%
  \BibitemOpen
  \bibfield  {author} {\bibinfo {author} {\bibfnamefont {G.}~\bibnamefont
  {Bir}}\ and\ \bibinfo {author} {\bibfnamefont {G.}~\bibnamefont {Pikus}},\
  }\href {https://books.google.ca/books?id=38m2QgAACAAJ} {\emph {\bibinfo
  {title} {{Symmetry and Strain-induced Effects in Semiconductors}}}}\
  (\bibinfo  {publisher} {Wiley},\ \bibinfo {address} {New York},\ \bibinfo
  {year} {1974})\BibitemShut {NoStop}%
\end{thebibliography}
%

\end{document}